\documentclass[12pt,showpacs,aps]{revtex4}
\usepackage{graphicx}
\usepackage{amsmath}
\usepackage{amsfonts}
\usepackage{amssymb}
\usepackage{wasysym}
\usepackage{color}
\usepackage{float}

\newcommand{\Sigmabold}{\boldsymbol\Sigma}
\newcommand{\omegabold}{\boldsymbol\omega}

\begin{document}

\title{Development of high vorticity structures and geometrical properties of the vortex line representation}
\thanks{dmitrij@itp.ac.ru}
\author{D.S. Agafontsev$^{(a),(b)}$, E.A. Kuznetsov$^{(b),(c)}$ and A.A. Mailybaev$^{(d)}$}
\affiliation{$^{(a)}$ P. P. Shirshov Institute of Oceanology, Moscow, Russia\\
$^{(b)}$ Novosibirsk State University, Novosibirsk, Russia\\
$^{(c)}$ P.N. Lebedev Physical Institute, Moscow, Russia\\
$^{(d)}$ Instituto Nacional de Matem\'atica Pura e Aplicada -- IMPA, Rio de
Janeiro, Brazil}

\begin{abstract}
The incompressible three-dimensional Euler equations develop very thin pancake-like regions of increasing vorticity. 
These regions evolve with the scaling $\omega_{\max}\propto l^{-2/3}$ between the vorticity maximum and the pancake thickness, as was observed in the recent numerical experiments \textit{[D.S. Agafontsev et al, Phys. Fluids 27, 085102 (2015)]}. 
We study the process of pancakes' development in terms of the vortex line representation (VLR), which represents a partial integration of the Euler equations with respect to conservation of the Cauchy invariants and describes compressible dynamics of continuously distributed vortex lines. 
We present, for the first time, the numerical simulations of the VLR equations with high accuracy, which we perform in adaptive anisotropic grids of up to $1536^3$ nodes. 
With these simulations, we show that the vorticity growth is connected with the compressibility of the vortex lines and find geometric properties responsible for the observed scaling $\omega_{\max}\propto l^{-2/3}$. 
\end{abstract}

\pacs{47.27.Cn, 47.27.De, 47.27.ek}
\maketitle

%----------------------------------------------------------------------------
%----------------------------------------------------------------------------

\section{Introduction}
\label{Sec:Intro}

The mechanism of vorticity growth in the incompressible 3D Euler equations was intensively studied over the last decades because of its relation to a possible finite-time blowup and subsequent transition to turbulence. 
Several analytical blowup and no-blowup criteria were established; see the reviews in~\cite{chae2008incompressible} and~\cite{gibbon2008three}. 
The central result is the Beale--Kato--Majda theorem~\cite{beale1984remarks}, which states that at a singular point (if it exists) the time integral of maximum vorticity must explode. 
In parallel, a large effort was made with numerical analysis. 
One of the early numerical studies~\cite{brachet1992numerical} examined evolution of periodic flows starting from random initial conditions and the symmetric Taylor--Green vortex. 
In both cases, maximum of vorticity was growing nearly exponentially with time and the regions of high vorticity represented pancake-like structures (thin vortex sheets) compressing in the transversal direction. 
The tendency toward a vortex sheet should suppress three-dimensionality of the flow and, hence, formation of a finite-time singularity, since the dynamics within 2D Euler equations is known to be regular, see e.g.~\cite{majda2002vorticity,pumir1990collapsing,ohkitani2008geometrical}. 
Thus, further numerical studies were mainly focused on specific initial conditions providing enhanced vorticity growth, e.g., antiparallel or orthogonal vortices; we refer to~\cite{gibbon2008three} and~\cite{agafontsev2015} for a brief review and to~\cite{hou2009blow,bustamante2012interplay,kerr2013bounds,brenner2016potential} for examples of recent numerical works. 
Despite these efforts, the existence of a blowup (unless it is triggered by physical boundary~\cite{luo2013potentially}) remains a highly controversial question. 
%~\cite{wolibner1933theoreme,kato1967classical,yudovich1963non,majda2002vorticity,pumir1990collapsing,ohkitani2008geometrical}.

In our previous papers~\cite{agafontsev2015,agafontsev2016development,agafontsev2016asymptotic} we returned to the problem of vorticity growth from generic large-scale initial conditions.  
We carried out several simulations in anisotropic grids of up to $2048^{3}$ total number of nodes and observed in details evolution of high-vorticity regions. 
We confirmed that these regions represent pancake-like structures and found that the flow near the pancake is described locally by a novel exact self-similar solution of the Euler equations combining a shear flow with an asymmetric straining flow. 
The maximum vorticity growth $\omega_{\max}(t)\propto e^{\beta_{2}t}$ and the pancake compression in the transversal direction $l(t)\propto e^{-\beta_{1}t}$ are characterized by significantly different exponents $\beta_{2}/\beta_{1}\approx 2/3$, leading to the Kolmogorov-type scaling law
\begin{equation}\label{introEq}
\omega_{\max}(t)\propto l(t)^{-2/3} 
\end{equation}
observed numerically during the pancake evolution. 
On the other hand, the pancake model solution allows for an arbitrary scaling between the vorticity maximum and the pancake thickness, and our observation of the $2/3$-scaling remained unexplained. 
Note that, rewritten for the velocity variation, the relation~(\ref{introEq}) has the same form as the $1/3$-H{\"o}lder velocity continuity necessary for the energy cascade in developed turbulence~\cite{landau2013fluid,frisch1999turbulence}. 
The pancake structures emerge in increasing number with time and provide the leading contribution to the energy spectrum, where, for some initial conditions~\cite{agafontsev2015,agafontsev2016development}, we observed the gradual formation of the Kolmogorov spectrum, $E(k)\propto k^{-5/3}$, in a fully inviscid flow. 

In the present paper we study the pancake vorticity structures from the point of view of the vortex line representation (VLR). 
The VLR is the transformation from the Eulerian coordinates of the fluid to the Lagrangian markers of the vortex lines~\cite{kuznetsov1998hamiltonian}, which is compressible, so that its Jacobian may take arbitrary values. 
In gas dynamics, the similar in spirit transformation from the Eulerian to the Lagrangian coordinates of the flow can be used to completely characterize the breaking areas~\cite{shandarin1989large,arnol1992catastrophe}. 
For compressible flows, a vanishing Jacobian of this transformation corresponds to intersection of particle trajectories and emergence of a singularity for the spatial derivatives of the velocity and density of the fluid. 
This process is accompanied by formation of a pancake-like structure with very different characteristic spatial scales along the breaking direction and in the transverse plane; in acoustics and optics such structures are called caustics. 
For incompressible flows, the vorticity is expressed explicitly through the VLR, and, as was suggested in~\cite{kuznetsov2000hamiltonian,kuznetsov2000collapse}, its growth may be explained by the vanishing Jacobian of the VLR mapping in the similar way as for the compressible case. 

The VLR equations appear as partial integration of the Euler equations with respect to conservation of the Cauchy invariants.  
Hence, a numerical simulation of the VLR equations must conserve the Cauchy invariants with the round-off accuracy along the vortex line trajectories. 
This property of the VLR simulations may be very important in the sense of the accuracy and control of the 3D Euler simulations while approaching sharp gradients. 
Note that the integral representation of the Cauchy invariants is the Kelvin theorem, stating that the velocity circulation is conserved for any closed contour moving with the flow.

Motivated by these observations, we develop a new numerical method for the Euler equations in terms of the VLR, and perform high-resolution simulations for two initial flows. 
We find that the regions of high vorticity do indeed correspond to decreasing VLR Jacobian, with the vorticity growing inverse-proportionally to the Jacobian. 
The inverse of the VLR Jacobian has the meaning of density of the vortex lines, so that the vorticity grows proportionally to this density. 
Combining the simulations with the pancake model solution of~\cite{agafontsev2016asymptotic}, we identify specific geometric properties responsible for the $2/3$-scaling~(\ref{introEq}), which originate from the next-order corrections to the model solution. 
We argue that the discussed approach may be applicable for a larger class of the so-called ``frozen-in-fluid'' fields advected by incompressible fluids, for instance, the magnetic field in MHD~\cite{kuznetsov2004compressible}. 
Note that while the original ideas of the VLR were suggested about 20 years ago~\cite{kuznetsov1998hamiltonian}, in this paper we present their first numerical examination performed with high accuracy.

We start with a general introduction to the frozen-in-fluid fields and the VLR in Section~\ref{Sec:VLR}. 
Detailed description of our numerical methods is given in Section~\ref{Sec:NumMethods} and Appendix~\ref{Sec:App0}. 
Section~\ref{Sec:NumResults} examines the basic features of the high-vorticity/low-Jacobian structures. 
In Section~\ref{Sec:Pancake_model} we study various properties of the VLR mapping and check them against the pancake model solution of~\cite{agafontsev2016asymptotic}. 
Using these findings, in Section~\ref{Sec:VLR_mapping} we identify geometric properties responsible for the scaling law~(\ref{introEq}). 
We finish with the Conclusions. 
Appendix~\ref{Sec:AppA} contains initial conditions and Appendix~\ref{Sec:AppB} describes results for the second simulation. 

%----------------------------------------------------------------------------
%----------------------------------------------------------------------------

\section{Vortex line representation}
\label{Sec:VLR}

For the 3D Euler equations 
\begin{equation}\label{Euler1}
\frac{\partial \mathbf{v}}{\partial t}+(\mathbf{v}\cdot \nabla )\mathbf{v}=-\nabla p,\quad \mathrm{div}\,\mathbf{v}=0,
\end{equation}
describing dynamics of ideal incompressible fluid of unit density,
the vorticity $\omegabold=\nabla\times\mathbf{v}$ 
satisfies the Helmholtz vorticity equation, 
\begin{equation}\label{Euler2}
\frac{\partial\omegabold}{\partial t}=\nabla\times(\mathbf{v}\times\omegabold).
\end{equation}
This equation shows that the vorticity represents an example a frozen-in-fluid divergence-free field. 
Indeed, with the divergence-free conditions, $\mathrm{div}\,\mathbf{v}=0$ and $\mathrm{div}\,\omegabold=0$, Eq.~(\ref{Euler2}) can be rewritten as 
\begin{equation}\label{Euler1-B}
\frac{d\omegabold}{dt}=(\omegabold\cdot \nabla )\mathbf{v},
\end{equation}
where $d/dt=\partial /\partial t+\mathbf{v}\cdot \nabla$ is the material time derivative. 
It has the same form as the evolution equation for an infinitesimal vector $\mathbf{\delta x}$ between the two fluid particles,
\begin{equation}\label{Euler1-dr}
\frac{d\,\mathbf{\delta x}}{dt}=\mathbf{\delta v}=(\mathbf{\delta x}\cdot\nabla)\mathbf{v}.
\end{equation}
This means that the vector $\mathbf{\delta x}$ initially parallel to $\omegabold$ retains this property  at later times, i.e., fluid particles belonging initially to the same vortex line (a line tangent to $\omegabold$ at each point) move together with this line. 
One can say that the vortex lines are ``frozen'' into the fluid during its motion. 
Other classical examples of frozen-in-fluid fields are the magnetic field in the magnetohydrodynamics~\cite{landau2013electrodynamics,moffatt1978field} and the divorticity field for 2D Euler hydrodynamics~\cite{kuznetsov2007effects}.

In the present paper we focus on the behavior of vortex lines. 
As follows from Eq.~(\ref{Euler2}), a vortex line can only be changed by the velocity component $\mathbf{v}_n$ perpendicular to $\omegabold$. 
To clarify this property, we introduce a new type of trajectories given by the modified equations of motion 
\begin{equation}\label{trajectories}
{\frac{d\mathbf{x}}{dt}=\mathbf{v}_{n}(\mathbf{x},t)}
\end{equation}
with the initial condition 
\begin{equation}\label{trajectoriesIC}
\mathbf{x}|_{t=0}=\mathbf{a}.
\end{equation}
A solution $\mathbf{x}=\mathbf{x}(\mathbf{a},t)$ may be seen as describing the motion of vortex lines. 
In terms of this solution, Eq.~(\ref{Euler2}) admits explicit integration in the form~\cite{kuznetsov2000hamiltonian,kuznetsov2008mixed} 
\begin{equation}\label{Cauchy}
\omegabold(\mathbf{x},t)=\frac{{\widehat{\mathbf{J}}}\,\omegabold_{0}(\mathbf{a})}{J},\quad \widehat{\mathbf{J}}(\mathbf{a},t)=[J_{ij}(\mathbf{a},t)]=\left[\frac{\partial x_{i}}{\partial a_{j}}\right], \quad J=\det {\widehat{\mathbf{J}}},
\end{equation}
where $\omegabold_{0}(\mathbf{a})$ is the initial vorticity at $t=0$ and $\widehat{\mathbf{J}}$ is the Jacobi matrix of the mapping $\mathbf{x}=\mathbf{x}(\mathbf{a},t)$. 
Relation~(\ref{Cauchy}) is the result of a partial integration of the Euler equations with respect to conservation of the Cauchy invariants. 
For the given case, the Cauchy invariants coincide with the initial vorticity $\omegabold_{0}(\mathbf{a})$ and can be understood as the flux of the velocity circulation for an infinitesimal loop in the space of Lagrangian markers $\mathbf{a}$. 

According to Eq.~(\ref{trajectories}) and the Liouville theorem, the Jacobian evolves as 
\begin{equation}\label{Liouville}
\frac{dJ}{dt} \equiv \bigg(\frac{\partial}{\partial t} + \mathbf{v}_{n}\cdot \nabla\bigg) J = \mathrm{div}(\mathbf{v}_{n})\,J,
\end{equation}
with the initial condition $J|_{t=0}=1$. 
Hence, the inverse of the Jacobian, $n=1/J$, which has the meaning of density of the vortex lines, satisfies the continuity equation, 
$$
\frac{\partial n}{\partial t}+\mathrm{div}(n\mathbf{v}_{n})=0.
$$
Generally, $\mathrm{div}\,\mathbf{v}_{n}\neq 0$, and the density $n$ is not preserved, emphasizing the compressible character of the mapping $\mathbf{x}=\mathbf{x}(\mathbf{a},t)$, despite the flow incompressibility and the divergence-free feature of the field $\omegabold$ itself.
In a generic case, a sustained growth of vorticity should be related to simultaneous decrease of the Jacobian in the denominator of Eq.~(\ref{Cauchy}), i.e., growth of the density $n$, what may be seen as formation of high density of vortex lines. 

Let us assume that the points of the vorticity maximum and the Jacobian minimum coincide. 
The vorticity maximum satisfies the so-called vortex-stretching equation, 
\begin{equation}\label{omega-max-t}
\frac{d\omega_{\max}}{dt} = \omega_{\max}\tau_{i}\tau_{j}\frac{\partial v_{i}}{\partial x_{j}} \approx \omega_{\max}\frac{\partial v_{\tau}}{\partial x_{\tau}} = \omega_{\max}\,\mathrm{div}(\mathbf{v}_{\tau}) = -\omega_{\max}\,\mathrm{div}(\mathbf{v}_{n}),
\end{equation}
where all spatial derivatives are taken at the maximum point, $\boldsymbol{\tau}$ is the unit vector along the direction $x_{\tau}$ of the vorticity and $v_{\tau}$ is the velocity component parallel to the vorticity; summation is implied with respect to repeated indexes. 
In Eq.~(\ref{omega-max-t}) we additionally assumed that the vorticity direction does not change significantly near the vorticity maximum, what corresponds to the results of our numerical simulations. 
According to Eq.~(\ref{Liouville}), the Jacobian minimum satisfies 
\begin{equation}\label{J-min-t}
\frac{d J_{\min}}{dt} = J_{\min}\,\mathrm{div}(\mathbf{v}_{n}), 
\end{equation}
and we conclude that the vorticity maximum should evolve inverse-proportionally to the Jacobian minimum, 
\begin{equation}\label{omega-J}
\omega_{\max}(t)J_{\min}(t)\approx \mathrm{const}.
\end{equation}
A high-vorticity region for which the vorticity direction changes sharply with the coordinate may feature a different relation between the vorticity and the Jacobian, as in this case we cannot bring the unit vector $\boldsymbol{\tau}$ under the spatial derivative in Eq.~(\ref{omega-max-t}).

Equations~(\ref{trajectories})--(\ref{Cauchy}) together with the relation $\omegabold=\nabla\times\,\mathbf{v}$ are called the \textit{vortex line representation} (VLR), and form a complete system equivalent to the Euler equations~\cite{kuznetsov1998hamiltonian}. 
However, this system is written in the mixed Eulerian ($\mathbf{x}$-space) and Lagrangian ($\mathbf{a}$-space) variables. 
For numerical study, we now rewrite it using the Eulerian variables.

Let $\mathbf{a}=\mathbf{a}(\mathbf{x},t)$ be the mapping inverse to $\mathbf{x}=\mathbf{x}(\mathbf{a},t)$. 
As follows from~(\ref{trajectories})--(\ref{trajectoriesIC}), this mapping obeys 
\begin{equation}\label{a-t}
\frac{\partial\mathbf{a}}{\partial t}+(\mathbf{v}_{n}\cdot\nabla)\mathbf{a}=0.
\end{equation}
Equation~(\ref{Cauchy}) can be rewritten in the form~\cite{kuznetsov2008mixed} 
\begin{equation}\label{omega-r}
\omega_{i}(\mathbf{x},t)=\frac{1}{2}\varepsilon_{ijk}\,\varepsilon_{\alpha\beta\gamma}\,\omega_{0\alpha}(\mathbf{a})\,\frac{\partial a_{\beta}}{\partial x_{j}}\,\frac{\partial a_{\gamma}}{\partial x_{k}},
\end{equation}
where $\omegabold_{0}(\mathbf{a})=(\omega_{01},\omega_{02},\omega_{03})$ is the initial vorticity and $\varepsilon_{ijk}$ is the Levi-Civita symbol. 
The two equations~(\ref{a-t}) and~(\ref{omega-r}) together with the relations 
\begin{equation}\label{v-normal-v}
\mathbf{v} = \mathrm{rot}^{-1}\omegabold = -\Delta^{-1}\,(\nabla\times\omegabold), \quad 
\mathbf{v}_{n} = \mathbf{v} - \frac{(\mathbf{v}\cdot\omegabold)}{\omega^{2}}\omegabold 
\end{equation}
for the velocity and the normal velocity represent complete VLR system of equations written in the Eulerian coordinates $(\mathbf{x},t)$.

%----------------------------------------------------------------------------
%----------------------------------------------------------------------------

\section{Numerical experiments}
\label{Sec:NumMethods}

We solve the system~(\ref{a-t})--(\ref{v-normal-v}) numerically in the box $\mathbf{x}=(x_1,x_2,x_3)\in[-\pi ,\pi]^{3}$ with periodic boundary conditions using the Runge--Kutta forth-order pseudo-spectral method. 
The initial conditions include the mapping $\mathbf{a}(\mathbf{x},0)=\mathbf{x}$ and the initial vorticity $\omegabold_{0}(\mathbf{x})$; the latter is given analytically, what allows for an exact calculation of $\omegabold_{0}(\mathbf{a})$ for a ``shifted'' $\mathbf{a}$-grid in Eq.~(\ref{omega-r}). 
In our implementation, we solve the VLR equations rewritten for the periodic mapping 
\begin{equation}\label{b-functions}
\mathbf{b}(\mathbf{x},t)=\mathbf{a}(\mathbf{x},t)-\mathbf{x},
\end{equation}
because the original mapping $\mathbf{a}(\mathbf{x},t)$ is not periodic. 
The inverse of the curl operator in Eq.~(\ref{v-normal-v}), as well as all spatial derivatives, are calculated in the Fourier space. 
We use adaptive anisotropic rectangular grid proposed in~\cite{agafontsev2015}, which is uniform for each direction and adapted independently along each coordinate; the adaption comes from the analysis of the Fourier spectrum of the solution. 
We start simulations in cubic grid $192^{3}$, continue with the fixed grid when the total number of nodes reaches $1536^{3}$, and finally stop when the Fourier spectrum becomes sufficiently wide. 
More details of the numerical scheme together with the tests of its high accuracy are given in Appendix~\ref{Sec:App0}.

As initial vorticity, we choose two flows $I_1$ and $I_2$ studied in~\cite{agafontsev2015} and summarized in Tabs.~\ref{tab:M1} and~\ref{tab:M2} in Appendix~\ref{Sec:AppA} along with some simulation information. 
These flows represent a superposition of the shear flow $\omegabold_0 = (\sin x_{3},\cos x_{3},0)$ and a random truncated (up to second harmonics) periodic perturbation. 
We have two main reasons for this choice. 
First, for small and moderate perturbation, the pancake vorticity structures emerging in such flows are compressed mainly along the $x_{3}$-axis, what allows to resolve the anisotropy of the solution in an optimal way and reach significantly higher detalization along the pancake transversal direction. 
This is especially important since the mapping~(\ref{b-functions}) features significantly wider Fourier spectrum than the vorticity field and the VLR equations have cubic nonlinearity, so that their simulation finishes significantly earlier in time than the direct simulation. 
Note that for generic large-scale initial conditions the high-vorticity regions represent pancake-like structures developing universally regardless of the presence of the shear flow~\cite{agafontsev2016asymptotic,agafontsev2016development}. 
Second, we only consider initial flows without null points, $\|\omegabold_{0}(\mathbf{a})\| \ne 0$, since otherwise the VLR scheme in Eq.~(\ref{v-normal-v}) has a topological singularity at a null point of vorticity, see e.g.~\cite{kuznetsov2008mixed}, and we prefer to avoid this complexity. 
For small and moderate periodic perturbations, our choice of initial conditions satisfies this property. 

In the paper, we focus on the simulation of the initial condition $I_{2}$, where the pancake structure associated with the global vorticity maximum appears early and we observe its evolution for a longer time interval; analogous results for the initial flow $I_{1}$ are summarized in Appendix~\ref{Sec:AppB}. For $I_{2}$, the simulation reaches the final time $t=7.5$ in the grid $1458\times 648\times 3456$, with about five-fold increase in the global vorticity maximum and with about two orders of magnitude for the ratio of the lateral pancake dimension to its thickness. 

As we noted in the previous Section, the VLR equations represent partly integrated Euler equations with respect to conservation of the Cauchy invariants. 
Hence, a numerical simulation of Eqs.~(\ref{a-t})--(\ref{v-normal-v}) must conserve the Cauchy invariants with the round-off accuracy along the vortex line trajectories given by the markers $\mathbf{a}$. 
This can be seen from Eq.~(\ref{omega-r}), where the vorticity is calculated directly from the initial vorticity $\omegabold(\mathbf{a}_{0})$ (the Cauchy invariant) and the mapping $\mathbf{a}(\mathbf{x},t)$. 
The conservation of the Cauchy invariants may be very important in the sense of the accuracy and control of the 3D Euler simulations while approaching sharp gradients. 
The only accumulating numerical error in the VLR simulations may come from calculation of the trajectories in Eq.~(\ref{a-t}), which can be controlled by using better space and time resolution. 
Comparing simulations in different grids and with different time steps in Appendix~\ref{Sec:App0}, we ensure that the numerical errors in calculation of the VLR mapping are very small and do not affect our results. 
%This means that the circulation should be very well conserved for any closed contour moving with the flow. 

%----------------------------------------------------------------------------
%----------------------------------------------------------------------------

\section{High-vorticity and low-Jacobian structures}
\label{Sec:NumResults}

\begin{figure}[t]
\centering
\includegraphics[width=6.7cm]{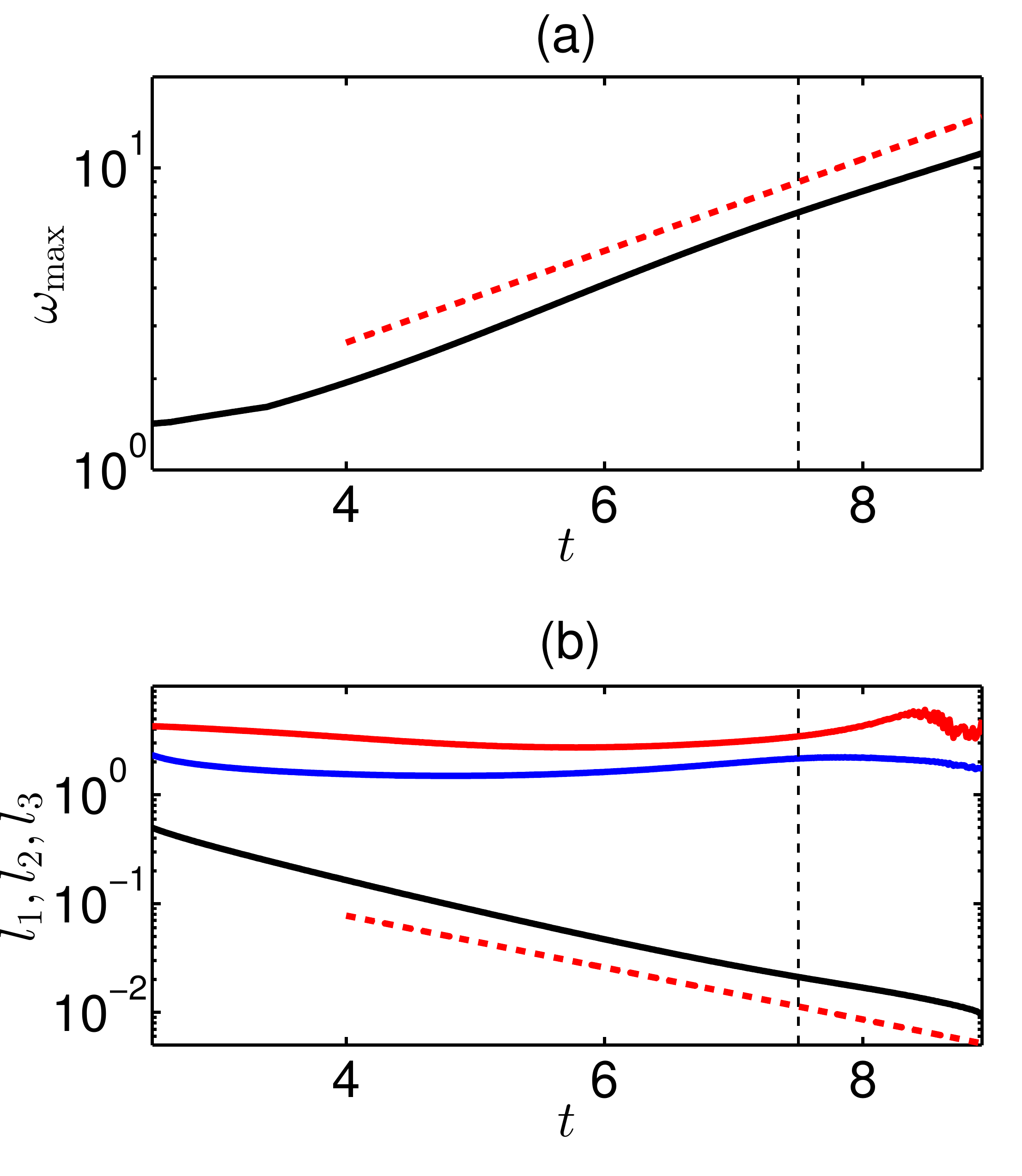}
\includegraphics[width=9.3cm]{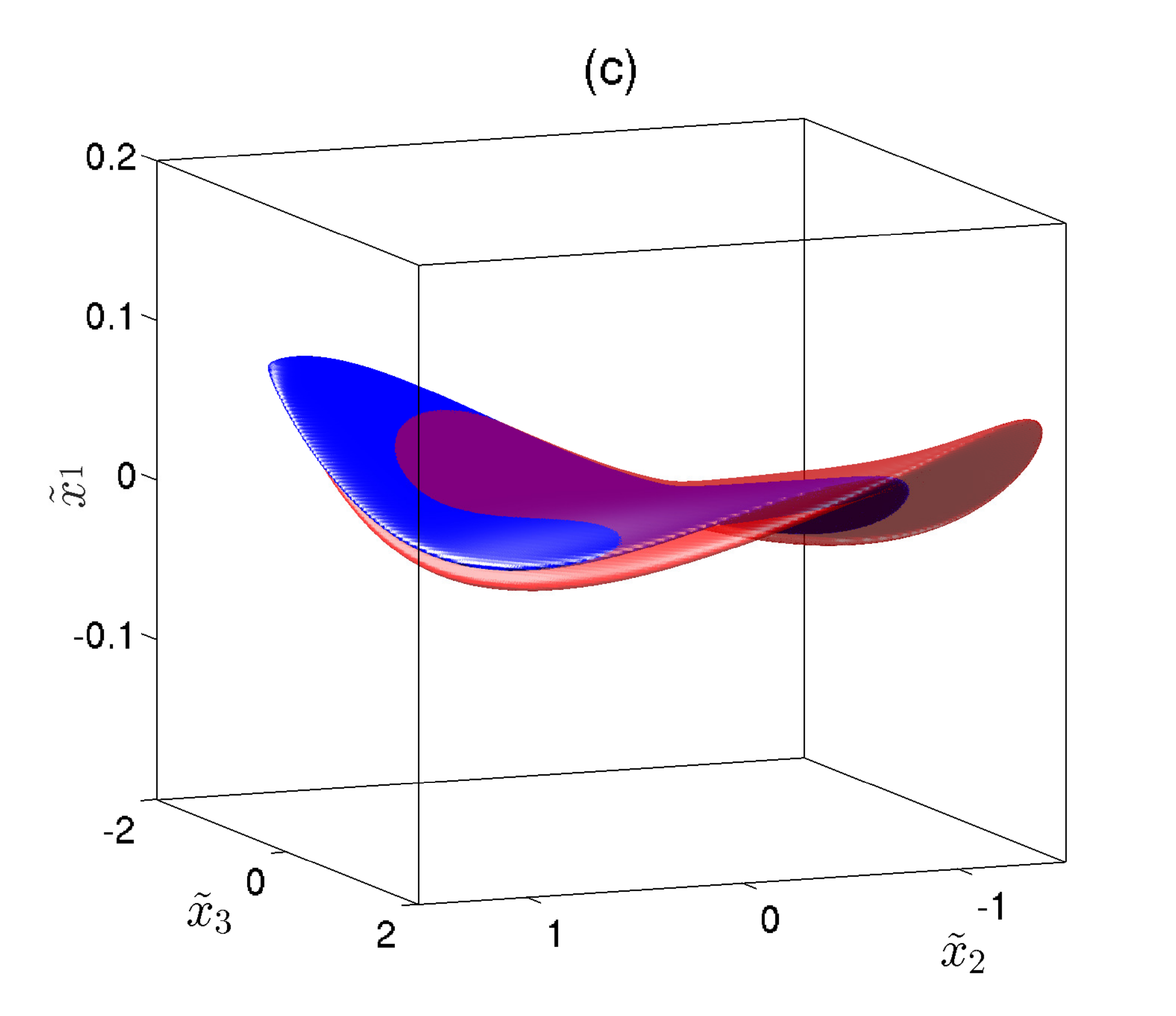}
\caption{\textit{(Color on-line)} 
(a) Global vorticity maximum as a function of time (logarithmic vertical scale) for the direct simulation of the Euler equations~(\ref{Euler2}). The red dashed line indicates the slope $\propto e^{\beta_{2}t}$ with $\beta_{2}=0.35$. The vertical dashed line marks the final time $t = 7.5$ for the simulation of the VLR equations~(\ref{a-t})--(\ref{v-normal-v}). 
(b) Time-evolution of the characteristic spatial scales $l_{1}$ (black), $l_{2}$ (blue) and $l_{3}$ (red) for the respective pancake vorticity structure. The red dashed line indicates the slope $\propto e^{-\beta_{1}t}$ with $\beta_{1}=0.55$. 
(c) Isosurfaces of vorticity $|\omegabold|=0.8\,\omega_{\max}$ (transparent red) and Jacobian $J=1.25\,J_{\min}$ (blue) in local $\tilde{\mathbf{x}}$-coordinates at $t=7.5$; the VLR simulation. Notice a much smaller vertical scale.}
\label{fig:fig1}
\end{figure}

To present our VLR results in a perspective, we remind our previous findings on the example of direct simulation of the Euler equations~(\ref{Euler2}), which we perform as described in~\cite{agafontsev2015,agafontsev2016asymptotic} for the $I_{2}$ initial flow in grid limited by $2048^3$ total number of nodes. 
The simulation reaches the final time $t=8.92$ in the grid $1944\times 972\times 4374$ with the global vorticity maximum $\omega_{\max}(t) = \max_{\mathbf{x}}|\omegabold(\mathbf{x},t)|$ increased by about $7.5$ times from its initial value $\omega_{\max}(0)=1.47$ to $11.2$. 
As shown in Fig.~\ref{fig:fig1}(a), for $t\ge 3.5$ this growth is nearly exponential, $\omega_{\max}(t)\propto e^{\beta_{2}t}$, with $\beta_{2}\approx 0.35$. 
The associated region of high vorticity becomes very thin at late times, see transparent red in Fig.~\ref{fig:fig1}(c). 
The figure uses the rotated coordinate axes aligned with the pancake geometry; the explicit definition of these coordinates is given in the end of this Section. 
The three characteristic sizes of the pancake can be estimated with the local second-order approximation as $l_i = \sqrt{2\,\omega_{\max}/|\lambda_i^{(\omega)}|}$, see~\cite{agafontsev2015}, where $|\lambda_1^{(\omega)}|\ge |\lambda_2^{(\omega)}|\ge |\lambda_3^{(\omega)}|$ are the three eigenvalues of the Hessian matrix $[\partial^2|\omegabold|/\partial x_{i}\partial x_{j}]$, computed at the point of maximum vorticity $\omega_{\max}$; the location of the maximum in between the grid nodes is approximated with the second-order finite-difference scheme. 
As illustrated in Fig.~\ref{fig:fig1}(b), the smallest size (thickness) $l_{1}$ decreases nearly exponentially with time, while the other two scales $l_{2}$ and $l_{3}$ do not change substantially, 
\begin{equation}
l_{1}\propto e^{-\beta_{1}t},\quad l_{2}\propto 1,\quad l_{3}\propto 1, \label{solution-exp-l}
\end{equation}
so that at the final time the span-to-thickness ratio reaches $l_2/l_1 \sim l_3/l_1 \gtrsim 100$. 
As one can see from the ratio of the exponents $\beta_{2}/\beta_{1}\approx 0.64$, the pancake follows the $2/3$-scaling law~(\ref{introEq}) between the vorticity maximum and the pancake thickness; we will return to this scaling in Section~\ref{Sec:VLR_mapping}. 

The new results from the VLR simulation confirm our previous conclusions. 
The end of the VLR simulation is marked by the vertical line in Fig.~\ref{fig:fig1}(a,b) at $t=7.5$, and until this line the vorticity fields for the two simulations practically coincide with each other, see Appendix~\ref{Sec:App0}. 
In addition to confirmation of the direct simulations, the VLR experiments allow accessing the properties of the VLR mapping $\mathbf{a}(\mathbf{x},t)$, what brings us new information. 

\begin{figure}
\centering
\includegraphics[width=7cm]{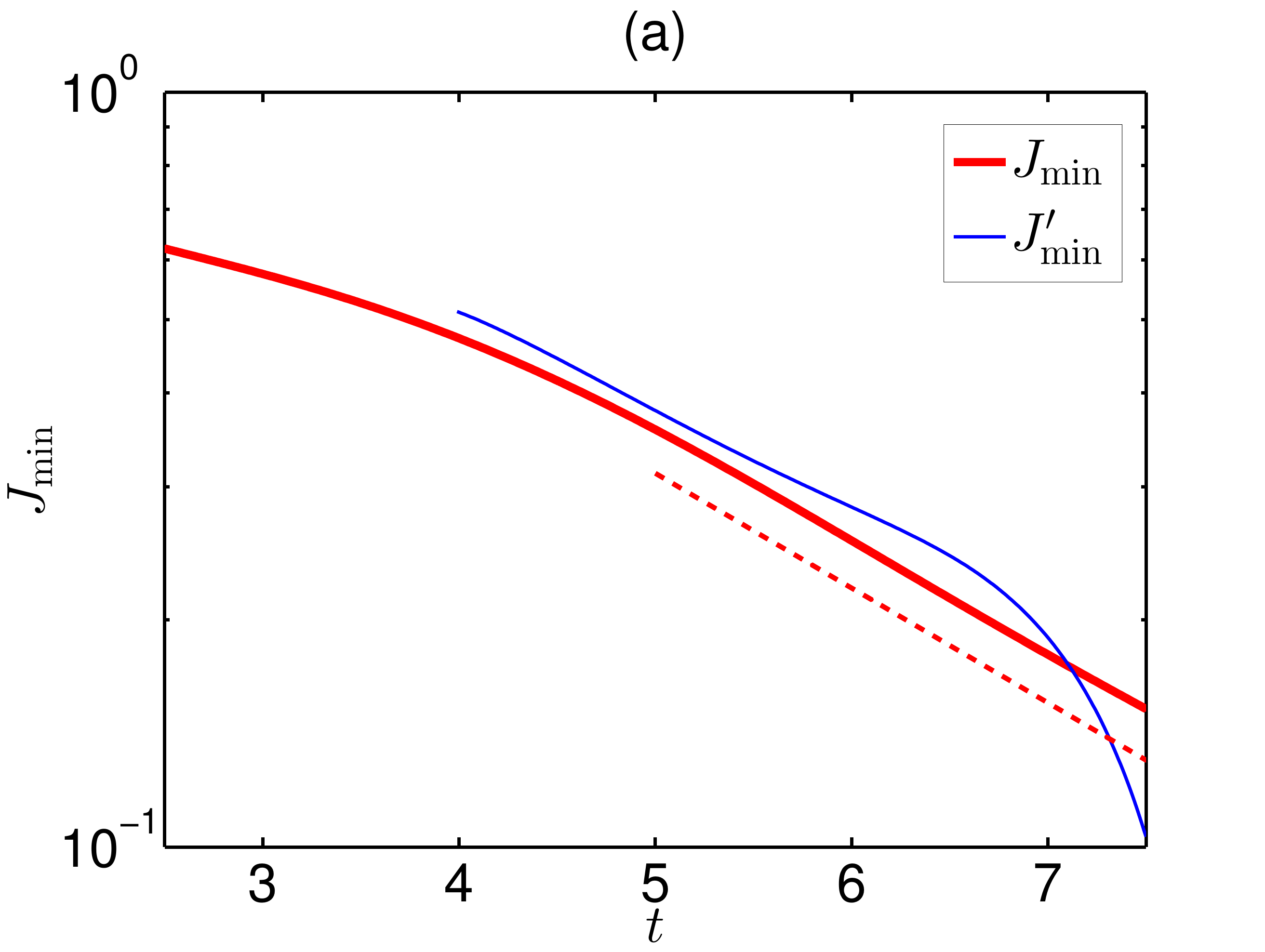}
\includegraphics[width=7cm]{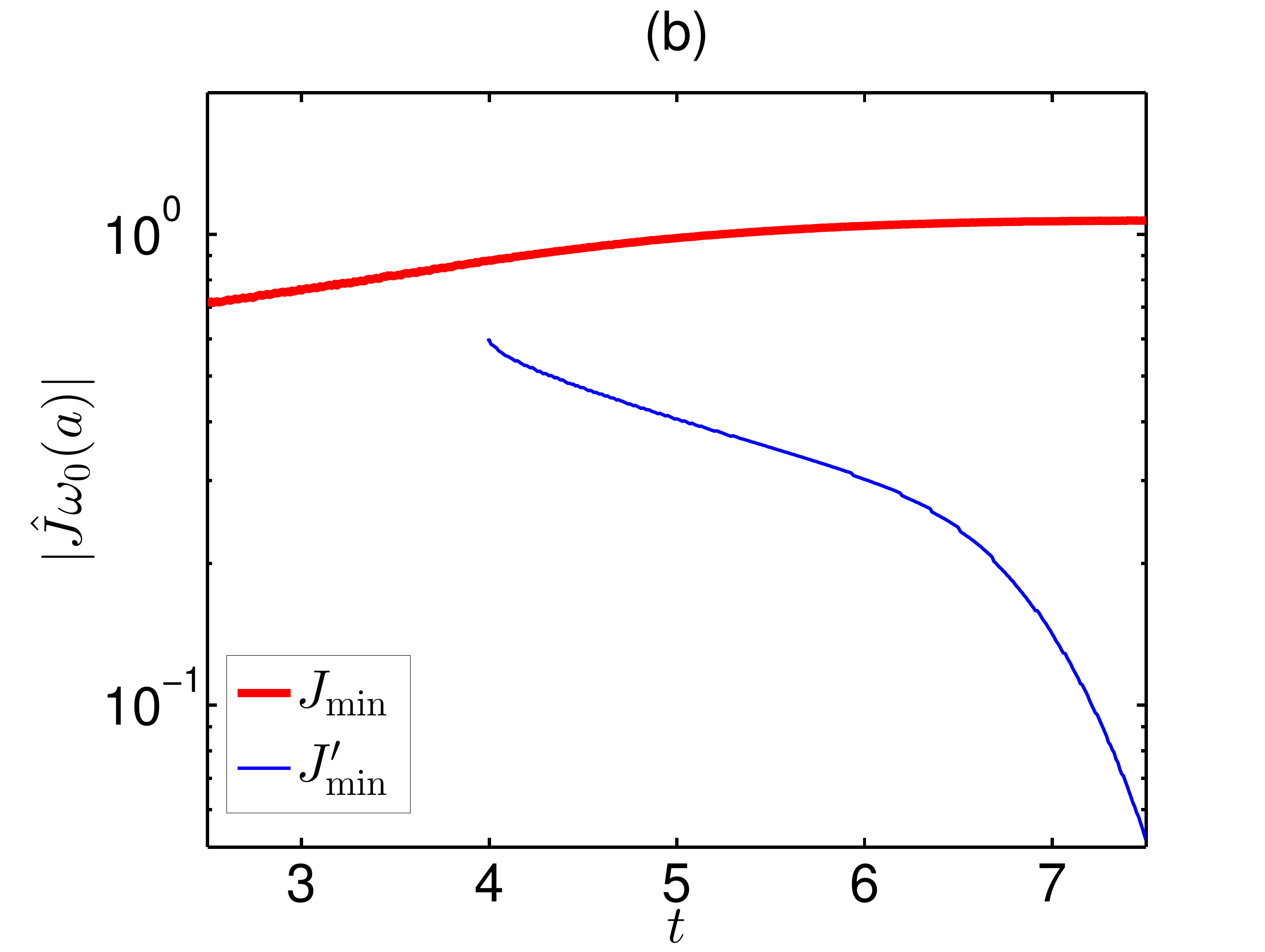}
\includegraphics[width=7cm]{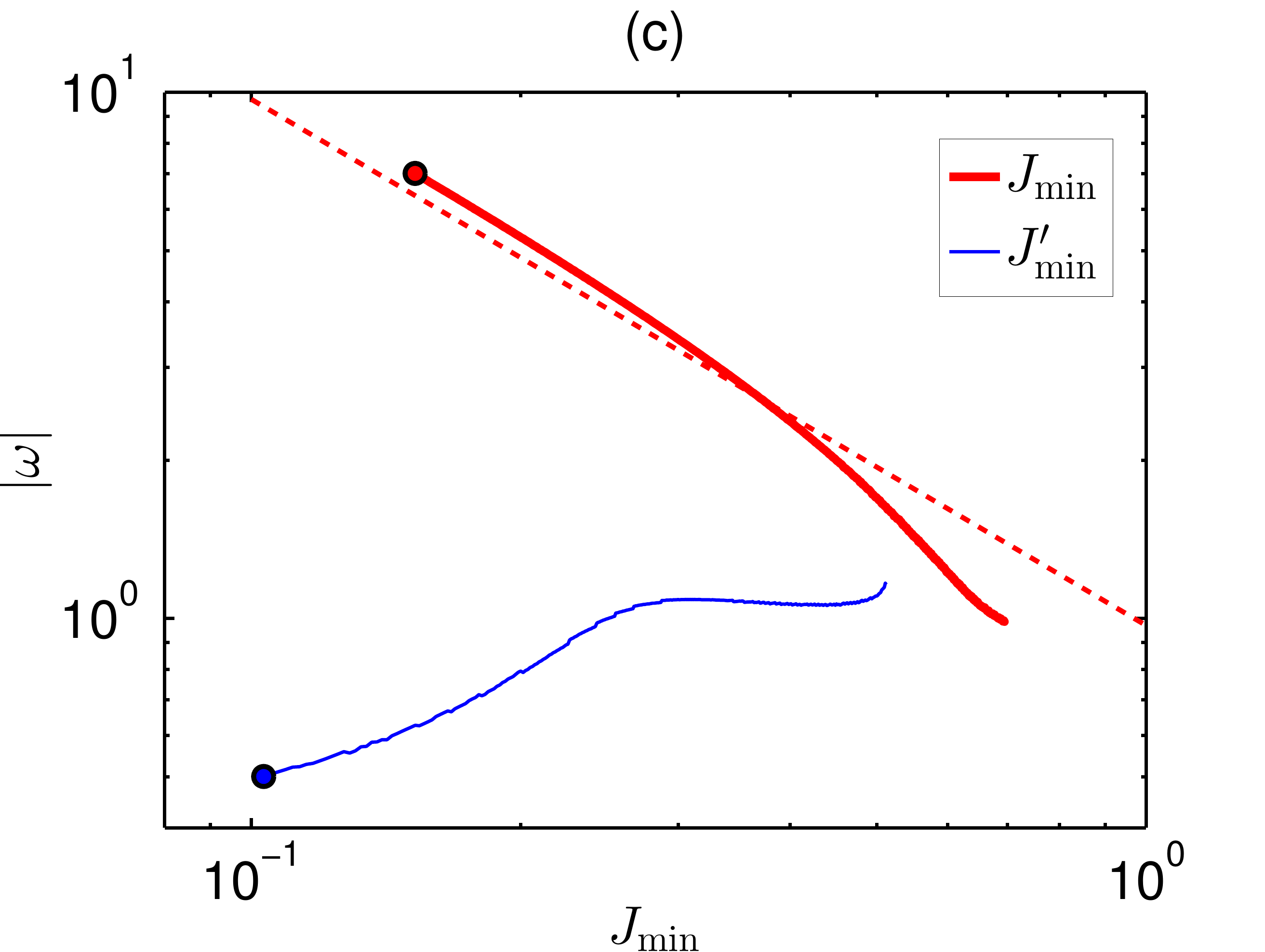}
\includegraphics[width=7cm]{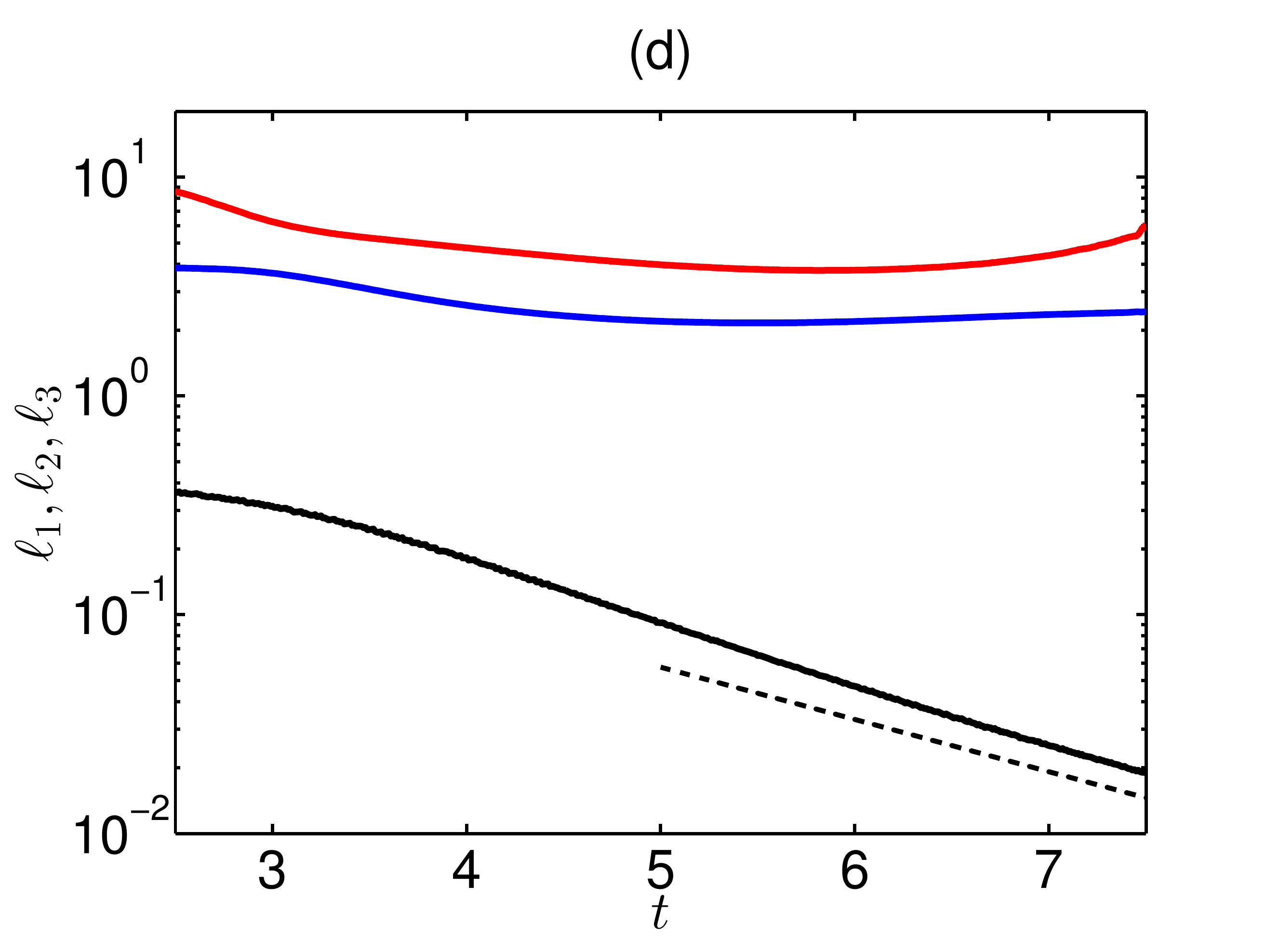}

\caption{\textit{(Color on-line)} 
(a) Time evolution of the two local Jacobian minima. The thick red curve represents the minimum $J_{\min}$ closest to the global vorticity maximum, and the thin blue curve corresponds to the ``parasitic'' minimum $J'_{\min}$. The red dashed line indicates the slope $\propto e^{-\beta_{2}t}$ with $\beta_{2}=0.35$. 
(b) Norms of the nominator ${\widehat{\mathbf{J}}}\,\omegabold_{0}(\mathbf{a})$ from Eq.~(\ref{Cauchy}) at the same points. 
(c) Relation between the vorticity and the Jacobian at the same points. The red dashed line indicates the asymptotic relation $\omega \propto 1/J_{\min}$, while the circles mark the final time $t=7.5$. 
(d) Time-evolution of the characteristic spatial scales $\ell_{1}$ (black), $\ell_{2}$ (blue) and $\ell_{3}$ (red) for the low-Jacobian structure corresponding to the minimum $J_{\min}$. The black dashed line indicates the slope $\propto e^{-\beta_{1}t}$ with $\beta_{1}=0.55$. 
}
\label{fig:fig2}
\end{figure}

In particular, as expected from the first relation in~(\ref{Cauchy}), the regions of large vorticity $|\omegabold|$ should correlate with the regions of small Jacobian $J$. 
This is confirmed in Fig.~\ref{fig:fig1}(c), where at the final time $t=7.5$ the (red) region of large vorticity around the global vorticity maximum mostly overlaps with the (blue) region of small Jacobian. 
The low-Jacobian region is associated with the Jacobian minimum $J_{\min}$, whose position is close to the vorticity maximum.
As shown by the time dependency in Fig.~\ref{fig:fig2}(a), $J_{\min}$ is the global minimum of the Jacobian until $t=7.1$ and becomes a local minimum for later times due to competition with another local minimum $J'_{\min}$.
Expression~(\ref{Cauchy}) relating the VLR with the vorticity contains the nominator ${\widehat{\mathbf{J}}}\,\omegabold_{0}(\mathbf{a})$; its norm and the resulting relation between the vorticity and the Jacobian are shown for both minima $J_{\min}$ and $J'_{\min}$ in Fig.~\ref{fig:fig2}(b,c). 
We see that  $J'_{\min}$ represents a ``parasitic'' minimum, when both the nominator and denominator decrease sharply, with no increase for the vorticity. 
Such a behavior is the result of alignment of the vector $\omegabold_{0}$ with the eigenvector corresponding to small eigenvalue of the Jacobi matrix $\widehat{\mathbf{J}}$; the angle between the two decreases to $14^{\circ}$ at the final time. 
We will disregard the minimum $J'_{\min}$ from now on, and focus on the study of the minimum $J_{\min}$. 

Figure~\ref{fig:fig2}(a) shows that the time dependency of the Jacobian minimum is close to exponential at late times, $J_{\min}\propto e^{-\beta_{2}t}$. 
This behavior is observed in approximately the same time interval and with the same exponent $\beta_{2}=0.35$ as for the vorticity maximum in Fig.~\ref{fig:fig1}(a). 
The latter leads to inverse-proportional relation 
\begin{equation}\label{vorticity-maximal}
\omega_{\max}(t)\propto\frac{1}{J_{\min}(t)}
\end{equation}
between the vorticity maximum and the Jacobian minimum, see the red curve in Fig.~\ref{fig:fig2}(c). 
Recalling the original exact equation~(\ref{Cauchy}), we conclude that the growth of the vorticity is related to the smallness of the Jacobian in the denominator, while the nominator plays a secondary role, as was suggested earlier in~\cite{kuznetsov2000hamiltonian,kuznetsov2000collapse,kuznetsov2008mixed} and we now confirm numerically in Fig.~\ref{fig:fig2}(b). 
Thus, the vorticity turns out to be proportional to the density $n=1/J$ of the vortex lines, in accordance with the relation~(\ref{omega-J}). 
We remind that we derived this relation under the assumptions that the positions of the vorticity maximum and the Jacobian minimum coincide and the vorticity direction does not change substantially near the vorticity maximum. 
We should note that the points of maximum vorticity and minimum Jacobian belong to the same pancake structure and stay close, but may not approach each other with time, as the distance between them decreases to $0.1$ at $t=5.9$ and then increases up to $0.44$ at the final time.

Geometry of the low-Jacobian structure in the $\mathbf{x}$-space, see the blue region in Fig.~\ref{fig:fig1}(c), can be described with the three characteristic scales $\ell_{i} = \sqrt{2J_{\min}/{\lambda}_{i}^{(J)}}$ in the same way as for the high-vorticity structure previously in this Section. 
Here ${\lambda}_{1}^{(J)}\ge {\lambda}_{2}^{(J)}\ge {\lambda}_{3}^{(J)}$ are the eigenvalues of the Hessian matrix $\boldsymbol{\Gamma}^{(x)} = [\partial^2J/\partial x_{i}\partial x_{j}]$ computed at the Jacobian minimum $J_{\min}$. 
These eigenvalues correspond to three orthonormal eigenvectors $\mathbf{n}_{1}$, $\mathbf{n}_{2}$ and $\mathbf{n}_{3}$, defining the local coordinate system $\tilde{\mathbf{x}}=\mathbf{n}_{1}\tilde{x}_{1}+\mathbf{n}_{2}\tilde{x}_{2}+\mathbf{n}_{3}\tilde{x}_{3}$ of the pancake used in Fig.~\ref{fig:fig1}(c). 
The three scales $\ell_{i}$ demonstrate the similar dynamics to that of the high-vorticity region, as can be seen from comparison of Fig.~\ref{fig:fig2}(d) and Fig.~\ref{fig:fig1}(b), i.e., 
\begin{equation}
\ell_{1}\propto e^{-\beta_{1}t},\quad \ell_{2}\propto 1,\quad \ell_{3}\propto 1. \label{solution-exp}
\end{equation}
Here the smallest scale $\ell_{1}$ (thickness) decreases nearly exponentially with approximately the same exponent $\beta_{1}=0.55$ as for the vorticity pancake thickness, $\ell_{1}\propto l_{1}$, while the other two scales do not change substantially. 

Note that, as we define the VLR trajectories with the initial conditions~(\ref{trajectoriesIC}), the transformation $\mathbf{x}=\mathbf{R}\,\tilde{\mathbf{x}}+\mathbf{x}_{m}$ to the local coordinate system of the pancake, where $\mathbf{R}=\{\mathbf{n}_{1},\mathbf{n}_{2},\mathbf{n}_{3}\}$ is the rotation matrix and $\mathbf{x}_{m}$ is the location of the Jacobian minimum, leads to the corresponding transformation $\mathbf{a}=\mathbf{R}\,\tilde{\mathbf{a}}+\mathbf{a}_{m}$, $\mathbf{a}_{m}=\mathbf{x}_{m}$, for the $\mathbf{a}$-variables. 
Hence, the Jacobi matrix $\tilde{\mathbf{J}}=[\partial\tilde{x}_{i}/\partial\tilde{a}_{j}]$ in the pancake coordinate system is related with the Jacobi matrix $\mathbf{J}=[\partial x_{i}/\partial a_{j}]$ in the original system as $\tilde{\mathbf{J}}=\mathbf{R}^{-1}\mathbf{J}\mathbf{R}$. 
All the results below are given for the rotated coordinate systems $\tilde{\mathbf{x}}$ and $\tilde{\mathbf{a}}$; we will omit the tildes for simplicity.

%----------------------------------------------------------------------------
%----------------------------------------------------------------------------

\section{Properties of the VLR mapping and the pancake model solution}
\label{Sec:Pancake_model}

In this Section, we study various properties of the VLR mapping for the high-vorticity/low-Jacobian structures, which we will use in Section~\ref{Sec:VLR_mapping} to examine the origin of the $2/3$-scaling~(\ref{introEq}). 
We check these properties against the analytical model for the pancake structure suggested in~\cite{agafontsev2016asymptotic} and show that most but not all of them are explained by this model. 
The model represents an exact self-similar solution of the Euler equations~(\ref{Euler2}) 
\begin{eqnarray}
\mathbf{v}(\mathbf{x},t) &=& -\omega_{\max}(t)\,l_{1}(t)\,f\left(\frac{x_1}{l_{1}(t)}\right) \mathbf{n}_{3} + \left(\begin{array}{c} -\beta_{1}x_{1} \\ \beta_{2}x_{2} \\ \beta_{3}x_{3} \end{array}\right), \label{solution-1}\\
\omegabold(\mathbf{x},t) &=& \omega_{\max}(t)f^{\prime}\left(\frac{x_1}{l_{1}(t)}\right) \mathbf{n}_{2},\label{solution-2}
\end{eqnarray}
written in Cartesian coordinates $\mathbf{x} = x_{1}\mathbf{n}_{1} + x_{2}\mathbf{n}_{2} + x_{3}\mathbf{n}_{3}$ and characterized with arbitrary constants $\beta_1$, $\beta_2$ and $\beta_3$ such that $-\beta_1 + \beta_2 + \beta_3 = 0$.
Here 
\begin{eqnarray}
\omega_{\max}(t)  = w_0 e^{\beta_{2}t},\quad
l_{1}(t) = h_0 e^{-\beta_{1}t},
\label{solution-2b}
\end{eqnarray}
are time dependencies for the vorticity maximum and the pancake thickness, $w_0$ and $h_0$ are positive prefactors and $f(\xi)$ is an arbitrary smooth function with $|\max f'(\xi)| = |f'(0)| = 1$.
The model~(\ref{solution-1})-(\ref{solution-2b}) describes a pancake of thickness $l_{1}$ oriented perpendicular to $x_{1}$-axis with the vorticity parallel to $x_{2}$-axis, and allows for any power-law dependency $\omega_{\max}\propto l_{1}^{-\beta_{2}/\beta_{1}}$ given by the ratio of the exponents $\beta_{2}/\beta_{1}$. 
Note that the numerical values of the exponents in our simulation are $\beta_{1}=0.55$, $\beta_{2}=0.35$ and $\beta_{3}=0.2$.

To construct the VLR, we calculate from~(\ref{solution-1})-(\ref{solution-2}) the velocity component normal to the vorticity,
\begin{equation}\label{solution-normal-v}
\mathbf{v}_{n}(\mathbf{x},t) = -\omega_{\max}(t)\,l_{1}(t)\,f\left(\frac{x_1}{l_{1}(t)}\right) \mathbf{n}_{3} + \left(\begin{array}{c} -\beta_{1}x_{1} \\ 0 \\ \beta_{3}x_{3} \end{array}\right).
\end{equation}
Then we solve~(\ref{trajectories}) with initial conditions~(\ref{trajectoriesIC}) and find the VLR mapping 
\begin{equation}\label{solution-mapping}
x_{1} = a_{1}\,e^{-\beta_{1}t},\quad x_{2} = a_{2},\quad 
x_{3} = a_{3}\,e^{\beta_{3}t} - w_0 h_0 f\left(\frac{a_{1}}{h_0}\right)\,\frac{\sinh(\beta_{3}t)}{\beta_{3}},
\end{equation}
where we used relation $\beta_1 = \beta_2+\beta_3$.
The corresponding Jacobi matrix and its determinant become
\begin{equation}\label{solution-Jacobi}
\widehat{\mathbf{J}}(\mathbf{a},t) 
= \bigg[\frac{\partial x_{i}}{\partial a_{j}}\bigg] = \begin{pmatrix} 
e^{-\beta_{1}t} & 0 & 0 \\ 0 & 1 & 0 \\ 
-w_0 f'\left(\frac{a_{1}}{h_0}\right)\,\frac{\sinh(\beta_{3}t)}{\beta_{3}} & 0 & e^{\beta_{3}t} 
\end{pmatrix},
\quad
J = \det \widehat{\mathbf{J}} = e^{(\beta_3-\beta_1) t} = e^{-\beta_2 t},
\end{equation}
and one can verify that expressions~(\ref{solution-2}),~(\ref{solution-mapping}) and~(\ref{solution-Jacobi})  satisfy the equation~(\ref{Cauchy}). 
Note that in terms of the VLR the pancake model solution is degenerate: the initial vorticity is aligned with the eigenvector of the Jacobi matrix corresponding to unit eigenvalue, ${\widehat{\mathbf{J}}}\,\omegabold_{0}=\omegabold_{0}$, so that all time-dependency for the vorticity comes from the denominator in relation~(\ref{Cauchy}), i.e., the Jacobian. 

As follows from~(\ref{solution-Jacobi}), the Jacobian $J = e^{-\beta_{2}t}$ is inverse-proportional to the maximum vorticity $\omega_{\max}\propto e^{\beta_{2}t}$ from (\ref{solution-2b}). 
This agrees with the numerical experiment, see Eq.~(\ref{vorticity-maximal}) and the graph in Fig.~\ref{fig:fig2}(c), and the relation~(\ref{omega-J}) derived under the assumption of unidirectional vorticity; the assumption is satisfied for the model~(\ref{solution-1})-(\ref{solution-2b}). 
Note, however, that the Jacobian in~(\ref{solution-Jacobi}) does not depend on spatial coordinates, while in simulations it changes sharply along the pancake perpendicular direction, see Fig.~\ref{fig:fig1}(c). 
We will examine this question in more details in the end of this Section. 

As we have demonstrated, increasing vorticity is related to decreasing Jacobian, and therefore the pancake vorticity structure should have a footprint in the VLR mapping $\mathbf{x}(\mathbf{a},t)$.
We can study the properties of the mapping using the singular-value decomposition (SVD) of the Jacobi matrix $\widehat{\mathbf{J}} = [\partial x_i/ \partial a_j]$ evaluated at the point of $J_{\min}$. 
The SVD represents the transformation $\widehat{\mathbf{J}} = \mathbf{U}\Sigmabold\mathbf{V}^T$ to a diagonal form $\Sigmabold=\mathrm{diag}\{\sigma_{1},\sigma_{2},\sigma_{3}\}$ containing real non-negative elements $0<\sigma_1<\sigma_2<\sigma_3$ called the singular values. 
Here $\mathbf{U}$ and $\mathbf{V}$ are real orthogonal matrices defining rotations in the $\mathbf{x}$- and $\mathbf{a}$-spaces respectively and $T$ stands for the matrix transpose. 
Thus, in the local bases induced by $\mathbf{U}$ and $\mathbf{V}$, the VLR mapping represents a stretching or compression along three orthogonal axes with the rates defined by the singular values $\sigma_1$, $\sigma_2$ and $\sigma_3$. 
The rotation matrices $\mathbf{U}$ and $\mathbf{V}$ can be computed as eigenvectors of symmetric matrices $\widehat{\mathbf{J}}\widehat{\mathbf{J}}^T$ and $\widehat{\mathbf{J}}^T\widehat{\mathbf{J}}$, respectively, while the singular values -- as square roots of eigenvalues of the same matrices. 
Note that $\mathbf{G}^{(a)}=\widehat{\mathbf{J}}^T\widehat{\mathbf{J}}$ is the metric tensor in the $\mathbf{a}$-space, 
\begin{equation}
G_{\alpha\beta}^{(a)} = \bigg[\frac{\partial x_{i}}{\partial a_{\alpha}}\frac{\partial x_{i}}{\partial a_{\beta}}\bigg],\quad d\mathbf{x}^{2} = G_{\alpha\beta}^{(a)}\,da_{\alpha}da_{\beta},  
\label{metric-tensor-1} 
\end{equation}
while $\mathbf{G}^{(x)}=[\widehat{\mathbf{J}}\widehat{\mathbf{J}}^T]^{-1}$ is the metric tensor in the $\mathbf{x}$-space, 
\begin{equation}
G_{ij}^{(x)} = \bigg[\frac{\partial a_{\alpha}}{\partial x_{i}}\frac{\partial a_{\alpha}}{\partial x_{j}}\bigg],\quad d\mathbf{a}^{2} = G_{ij}^{(x)}\,dx_{i}dx_{j}. 
\label{metric-tensor-2} 
\end{equation}

\begin{figure}[t]
\centering
\includegraphics[width=7.5cm]{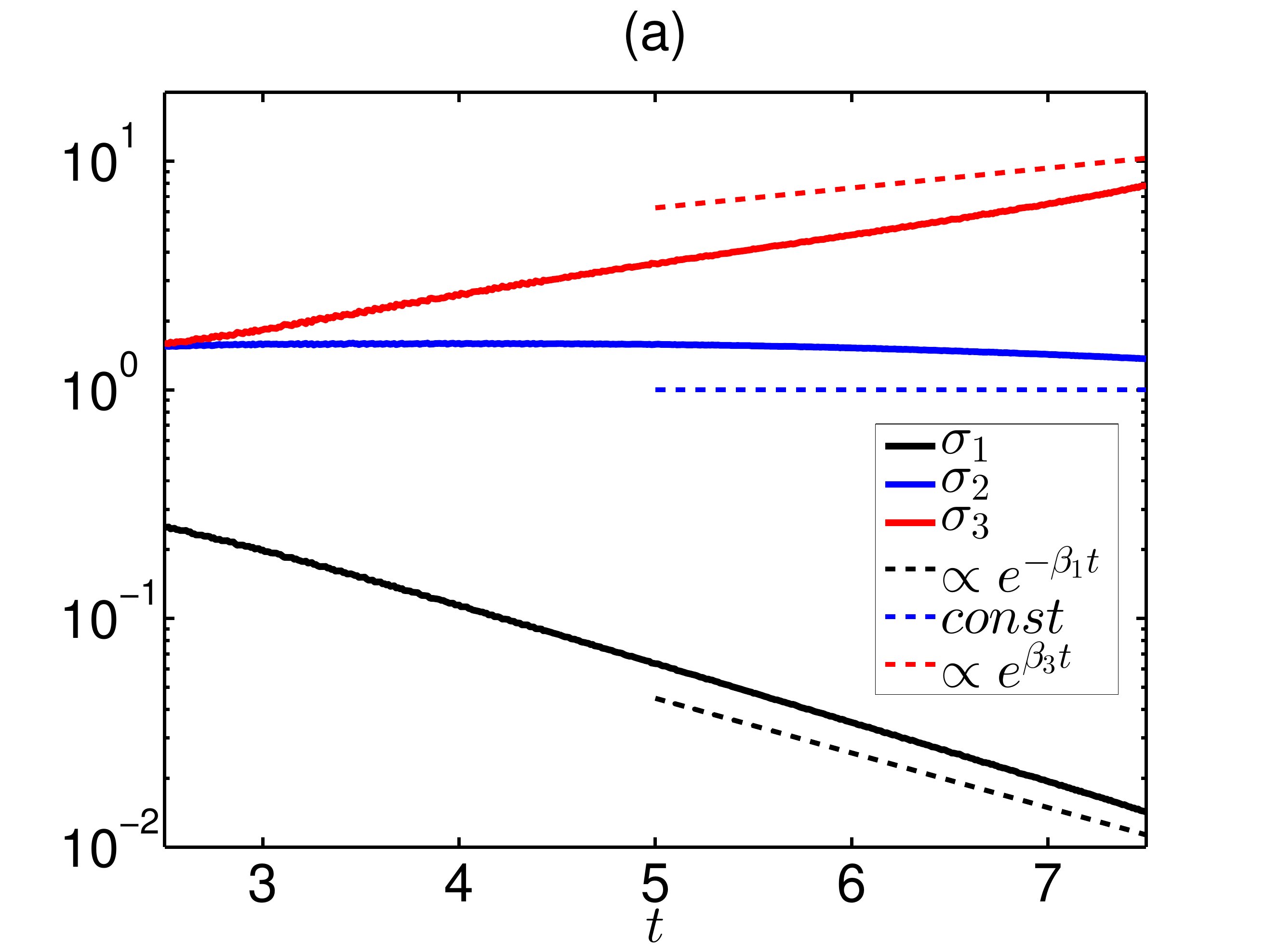}
\includegraphics[width=7.5cm]{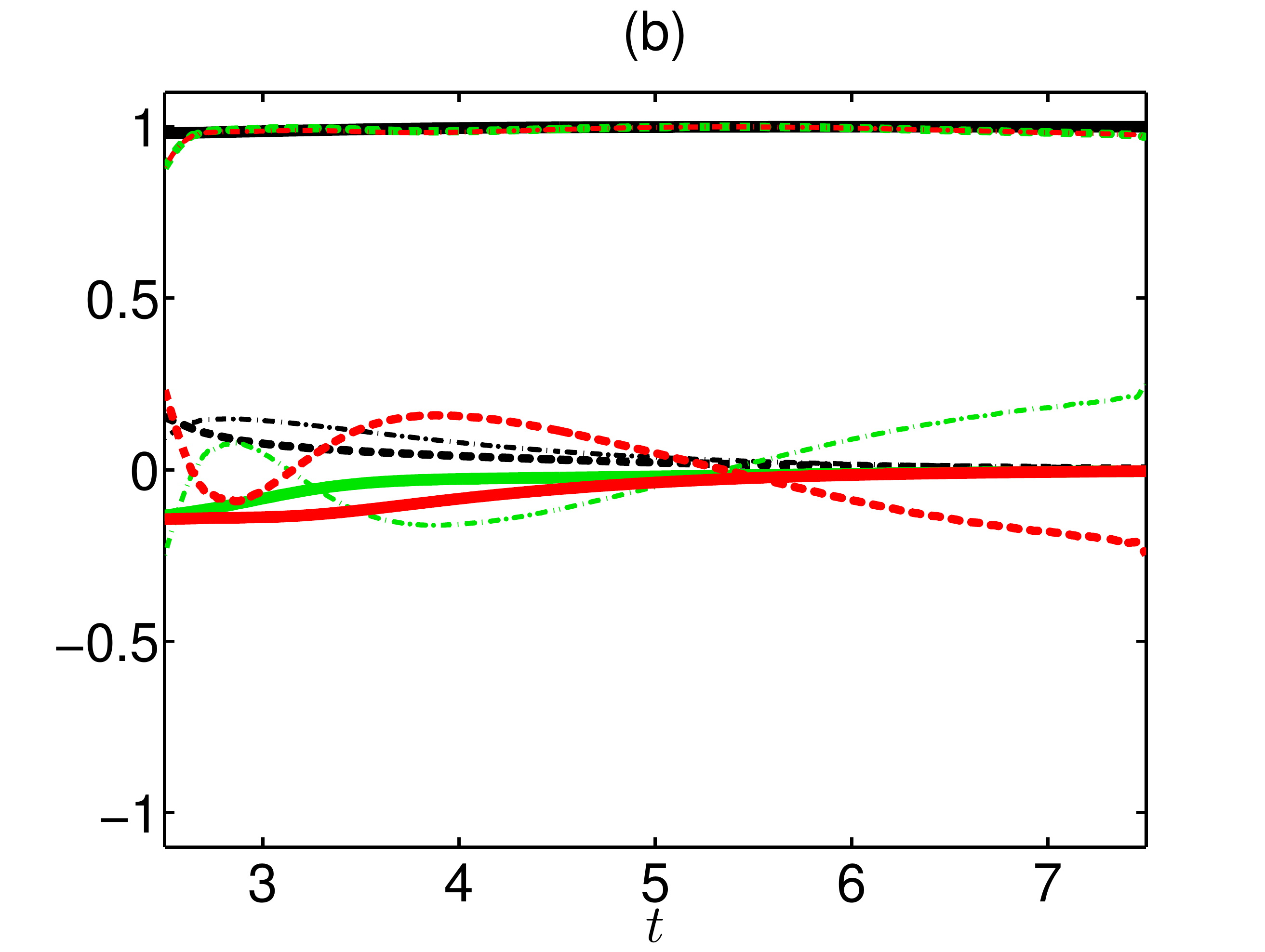}
\includegraphics[width=7.5cm]{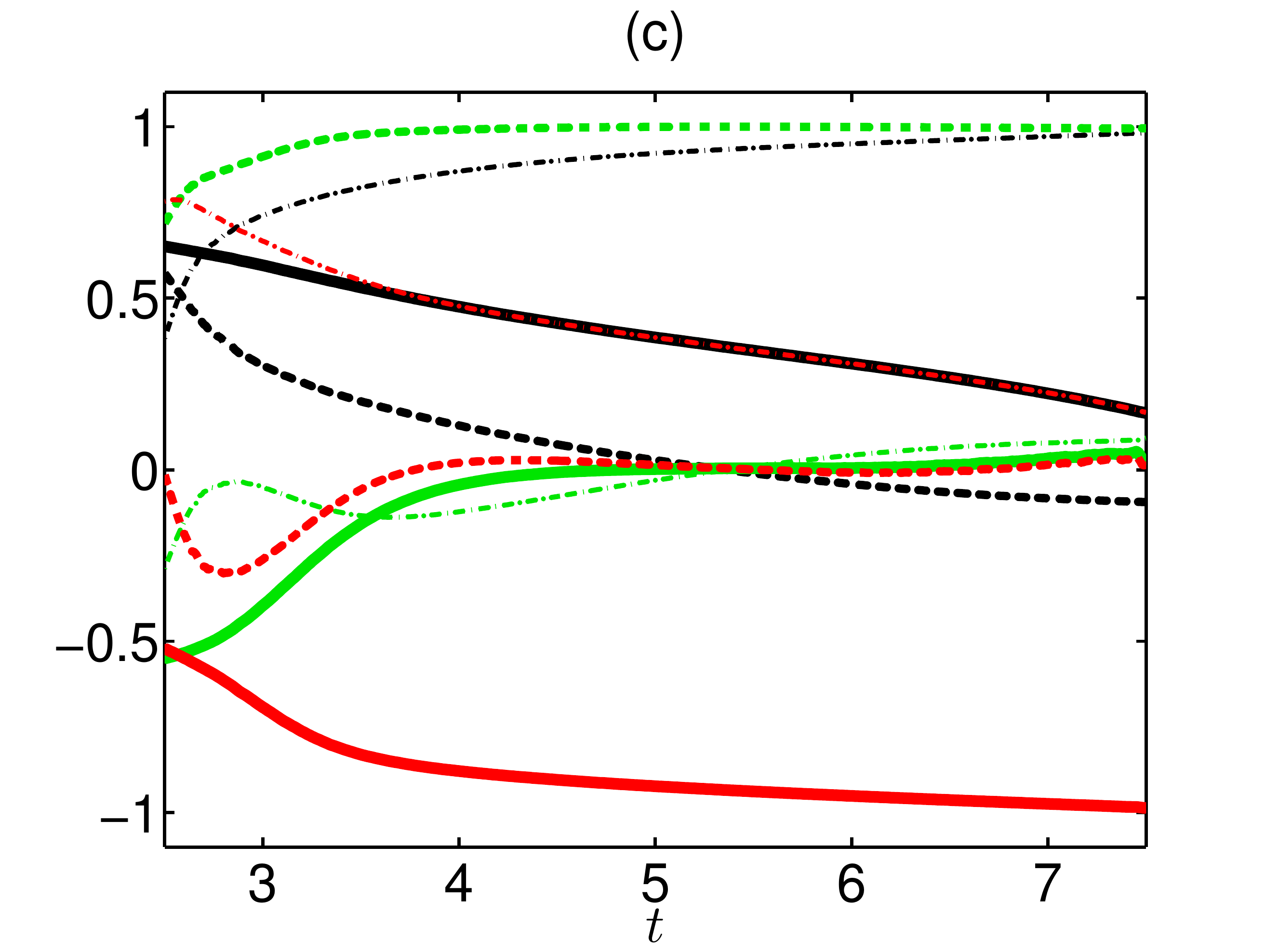}
\includegraphics[width=7.5cm]{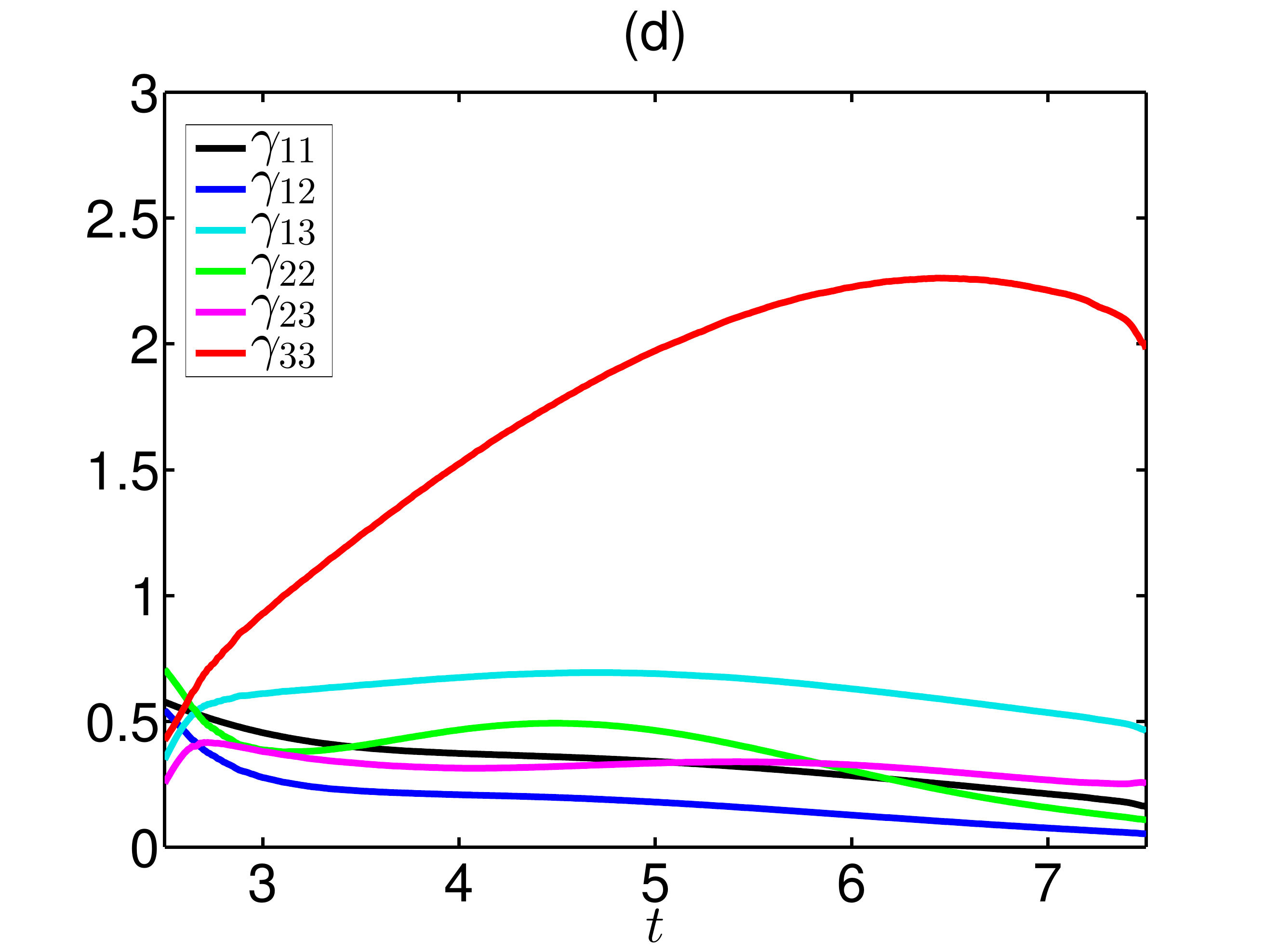}
\caption{
\textit{(Color on-line)} 
(a) Singular values $\sigma_{1}$ (black), $\sigma_{2}$ (blue) and $\sigma_{3}$ (red) of the Jacobi matrix $\widehat{\mathbf{J}}$ computed at $J_{\min}$, as functions of time. Dashed lines show the exponential slopes~(\ref{solution-S}).
(b) Components of the rotation matrix $\mathbf{U}$ in $\mathbf{x}$-space, as functions of time. Thick black curve shows the $(1,1)$ component, dashed black -- $(1,2)$, thin dash-dot black -- $(1,3)$, thick green -- $(2,1)$, dashed green -- $(2,2)$, thin dash-dot green -- $(2,3)$, thick red -- $(3,1)$, dashed red -- $(3,2)$, thin dash-dot red -- $(3,3)$. 
(c) Same for the rotation matrix $\mathbf{V}$ in $\mathbf{a}$-space. 
(d) Evolution of the Hessian $\boldsymbol{\gamma} = \mathbf{V}^{T}\boldsymbol{\Gamma}^{(a)}\mathbf{V}$ in $\hat{V}$-basis.  
}
\label{fig:fig3}
\end{figure}

For the pancake model solution, the Jacobi matrix~(\ref{solution-Jacobi}) has the singular values 
\begin{equation}
\sigma_1^2 = g-\sqrt{g^2-e^{-2\beta_2t}},\quad
\sigma_2^2 = 1, \quad 
\sigma_3^2 = g+\sqrt{g^2-e^{-2\beta_2t}}, 
\label{solution-S0} 
\end{equation}
where 
\begin{equation}
g = \frac{1}{2}\left(e^{-2\beta_{1}t}+e^{2\beta_{3}t}+
\left[w_0 f'\left(\frac{a_{1}}{h_0}\right)\,\frac{\sinh(\beta_{3}t)}{\beta_{3}}\right]^2
\right).
\label{solution-S0b} 
\end{equation}
The leading terms in the large-time asymptotic of~(\ref{solution-S0}) have the form
\begin{equation}
\sigma_1 \propto e^{-\beta_1t},\quad
\sigma_2 = 1, \quad 
\sigma_3 \propto e^{\beta_3t}.
\label{solution-S} 
\end{equation}
This agrees reasonably well with the numerical results in Fig.~\ref{fig:fig3}(a), where the exponents are presented by the dashed lines. 
Thus, the VLR mapping near the Jacobian minimum (and, therefore, in the high-vorticity pancake) is strongly compressed along one direction with the rate proportional to the pancake thickness, $\sigma_1\propto e^{-\beta_1t}\propto l_{1}$, and stretched along the other direction as $\sigma_3\propto e^{\beta_3t}\propto \omega_{\max}^{-1}l_{1}^{-1}$, while the remaining singular value $\sigma_2$ is close to unity and does not change with time significantly. 
Assuming that such behavior persists in the limit $t\to \infty$, the Lagrangian markers $\mathbf{a}$ distributed initially along the direction corresponding to the first singular value $\sigma_1$ collapse to a point. 
This may be seen as touching of the corresponding vortex lines, with the vorticity growing unboundly $\omega_{\max}(t)\to \infty$. 

The rotation matrices of the SVD provide the orthonormal bases $\mathbf{U} = \{\mathbf{n}^{(x)}_{1}, \mathbf{n}^{(x)}_{2}, \mathbf{n}^{(x)}_{3}\} $ and $\mathbf{V} = \{\mathbf{n}^{(a)}_{1}, \mathbf{n}^{(a)}_{2}, \mathbf{n}^{(a)}_{3}\}$ in the $\mathbf{x}$- and  $\mathbf{a}$-spaces. 
For the leading terms in the large-time asymptotic of $\mathbf{U}$ and $\mathbf{V}$, the pancake model solution yields 
\begin{eqnarray}
\mathbf{U} \simeq \mathbf{1}, \quad
\mathbf{V} \simeq \begin{pmatrix} \frac{1}{\sqrt{1+q^{2}}} & 0 & \frac{q}{\sqrt{1+q^{2}}} \\ 0 & 1 & 0 \\ \frac{-q}{\sqrt{1+q^{2}}} & 0 & \frac{1}{\sqrt{1+q^{2}}} \end{pmatrix}, \label{solution-UV} 
\end{eqnarray}
where 
\begin{equation}\label{solution-UVg}
q=-\frac{w_{0}}{2\beta_{3}} f'\bigg(\frac{a_{1}}{h_{0}}\bigg).
\end{equation}
This is supported by our numerical results, see Fig.~\ref{fig:fig3}(b,c): at late times both matrices do not change substantially, $\mathbf{U}$ is close to unity and $\mathbf{V}$ is close to anti-diagonal matrix with elements $V_{13}\approx V_{22}\approx -V_{31}\approx 1$.
Note that the ratio $w_{0}/2\beta_{3}$ in Eq.~(\ref{solution-UVg}) may take arbitrary values depending on the vorticity maximum $w_{0}$ at the initial time $t=0$, see Eq.~(\ref{solution-2b}), and the observed behavior for $\mathbf{V}$ can be obtained in the case of large $q$. 

The VLR mapping $\mathbf{x}(\mathbf{a},t)$ represents current positions $\mathbf{x}$ of markers $\mathbf{a}$ of the vortex lines, and its Jacobi matrix describes transformation of an infinitesimal vector $\mathbf{\delta a}$ with the vortex line motion, $\mathbf{\delta x} = \widehat{\mathbf{J}}\,\mathbf{\delta a}$. 
Substituting the SVD and solving the latter equality with respect to $\mathbf{\delta a}$, we get 
$$
\mathbf{\delta a} = \mathbf{V}\Sigmabold^{-1}\mathbf{U}^{T}\mathbf{\delta x},
$$
where we used orthogonality of the rotation matrices. 
Since $\mathbf{U} \simeq \mathbf{1}$, a vector $\mathbf{\delta a}_{n}$ that turns into the pancake perpendicular direction $\mathbf{\delta x}_{n} = (\delta x_{1},0,0)^{T}$ has length inverse-proportional to the first singular value, $|\mathbf{\delta a}_{n}|\propto \sigma_{1}^{-1}\delta x_{1}$ (the same result is obtained by using Eqs.~(\ref{solution-Jacobi}),~(\ref{solution-S0})-(\ref{solution-S}) for the pancake model solution). 
Hence, vortex lines within the pancake structure separated along the transversal direction by distance $\propto l_{1}$ in the $\mathbf{x}$-space, have distance between them in the $\mathbf{a}$-space $\propto l_{1}/\sigma_{1}\propto 1$, i.e., the pancake thickness in the $\mathbf{a}$-space remains of unity order. 
This conclusion differs from the previously made assumption~\cite{kuznetsov2000collapse} that, similarly to gas dynamics, the region of high vorticity shrinks in the space of Lagrangian markers $\mathbf{a}$ with time. 

As we have shown, most of the VLR mapping properties observed in the numerical simulations -- the inverse-proportionality between the vorticity maximum and the Jacobian minimum~(\ref{vorticity-maximal}) together with the time-dependencies for the singular values~(\ref{solution-S}) and the rotation matrices~(\ref{solution-UV}) -- are explained analytically by the pancake model solution suggested in~\cite{agafontsev2016asymptotic}. 
However, there is one particular property that is not explained -- this is the sharp dependency of the Jacobian along the pancake perpendicular direction $x_{1}$, see Fig.~\ref{fig:fig1}(c); we remind that for the pancake model the Jacobian~(\ref{solution-Jacobi}) does not depend on spatial coordinates. 
Let us assume, nonetheless, that the spatial dependency is present due to the next-order corrections. 
Then, as we demonstrate below, the observed behavior for the Jacobian along the $x_{1}$-axis comes from small but finite second derivatives of the Jacobian in the $\mathbf{a}$-space. 

At the Jacobian minimum, the Hessian matrices $\hat{\Gamma}^{(x)}=[\partial^{2}J/\partial x_{i}\partial x_{j}]$ and $\hat{\Gamma}^{(a)}=[\partial^{2}J/\partial a_{i}\partial a_{j}]$ are connected by the chain rule 
\begin{equation}\label{HessianJxa}
\boldsymbol{\Gamma}^{(a)} = \widehat{\mathbf{J}}^T \boldsymbol{\Gamma}^{(x)}\widehat{\mathbf{J}}.
\end{equation}
Note that calculation of the Hessian in the $\mathbf{a}$-space is a nontrivial problem, which we solve with the relations 
\begin{equation}\label{HessianJxa-xa}
\frac{\partial}{\partial a_{i}} = \frac{\partial x_{\alpha}}{\partial a_{i}}\frac{\partial}{\partial x_{\alpha}},\quad
\frac{\partial x_{\alpha}}{\partial a_{i}} = \frac{1}{2J}\epsilon_{ijk}\epsilon_{\alpha\beta\gamma}\frac{\partial a_{j}}{\partial x_{\beta}}\frac{\partial a_{k}}{\partial x_{\gamma}}. 
\end{equation}
Substituting the SVD $\widehat{\mathbf{J}} = \mathbf{U}\Sigmabold\mathbf{V}^T$ into~(\ref{HessianJxa}) and using the orthogonality of the matrix $\mathbf{V}$, we obtain 
\begin{equation}\label{HessianJ_xa}
\boldsymbol{\gamma} = \mathbf{V}^{T}\boldsymbol{\Gamma}^{(a)}\mathbf{V} = \Sigmabold\mathbf{U}^{T}\boldsymbol{\Gamma}^{(x)}\mathbf{U}\Sigmabold,
\end{equation}
where $\boldsymbol{\gamma}$ is the Hessian in the basis induced by rotation matrix $\mathbf{V}$. 
Since in the pancake coordinate system, see Section~\ref{Sec:NumResults}, $\boldsymbol{\Gamma}^{(x)}$ is diagonal and $\mathbf{U}$ is close to unity, the Hessian $\boldsymbol{\gamma}$ must be close to a diagonal matrix too. 
This agrees with the numerical simulations, see Fig.~\ref{fig:fig3}(d), where at late times $\boldsymbol{\gamma}$ has only the $(3,3)$-component which is of unity order and does not change significantly, while the other components are small. 
As $\mathbf{U}$ is close to unity, the relation~(\ref{HessianJ_xa}) yields 
\begin{equation}\label{HessianJ_xaB}
\gamma_{ii} \approx \sigma_i^2\lambda_i^{(J)},\quad i =1,2,3,
\end{equation}
where $\lambda_i^{(J)}$ are the eigenvalues of the Hessian $\boldsymbol{\Gamma}^{(x)}$. 
Then, expanding in the $\mathbf{x}$-space the Jacobian near its minimum, we get 
\begin{equation}\label{Jexpansion_xa}
J-J_{\min}\approx \frac{1}{2}\sum_{i=1}^{3}\lambda^{(J)}_{i}\Delta x_{i}^{2}\approx \frac{1}{2}\sum_{i=1}^{3}\gamma_{ii}\bigg(\frac{\Delta x_{i}}{\sigma_{i}}\bigg)^{2}. 
\end{equation}
Since only the first singular value of the Jacobi matrix is small and exponentially decreases with time, $\sigma_{1}\propto l_{1}\propto e^{-\beta_{1}t}$, all terms except the one proportional to $\Delta x_{1}^{2}$ can be neglected, so that 
\begin{equation}\label{Jexpansion_xa2}
J-J_{\min}\approx \frac{\gamma_{11}}{2}\bigg(\frac{\Delta x_{1}}{\sigma_{1}}\bigg)^{2}\propto \gamma_{11}\bigg(\frac{\Delta x_{1}}{l_{1}}\bigg)^{2}.
\end{equation}
Hence, we conclude that the sharp dependency of the Jacobian in the $x_{1}$-direction comes from small but finite component $\gamma_{11}$. 

We stress that the pancake model solution has zeroth spatial second derivatives of the Jacobian. 
However, if we assume that these second derivatives are present due to the next-order corrections, and remain finite, the pancake model allows to explain both the diagonal form of the matrix $\boldsymbol{\gamma}$, see Eq.~(\ref{HessianJ_xa}), and the sharp dependency of the Jacobian along the pancake perpendicular direction. 

%----------------------------------------------------------------------------
%----------------------------------------------------------------------------

\section{Origin of the $2/3$-scaling}
\label{Sec:VLR_mapping}

\begin{figure}[t]
\centering
\includegraphics[width=7.5cm]{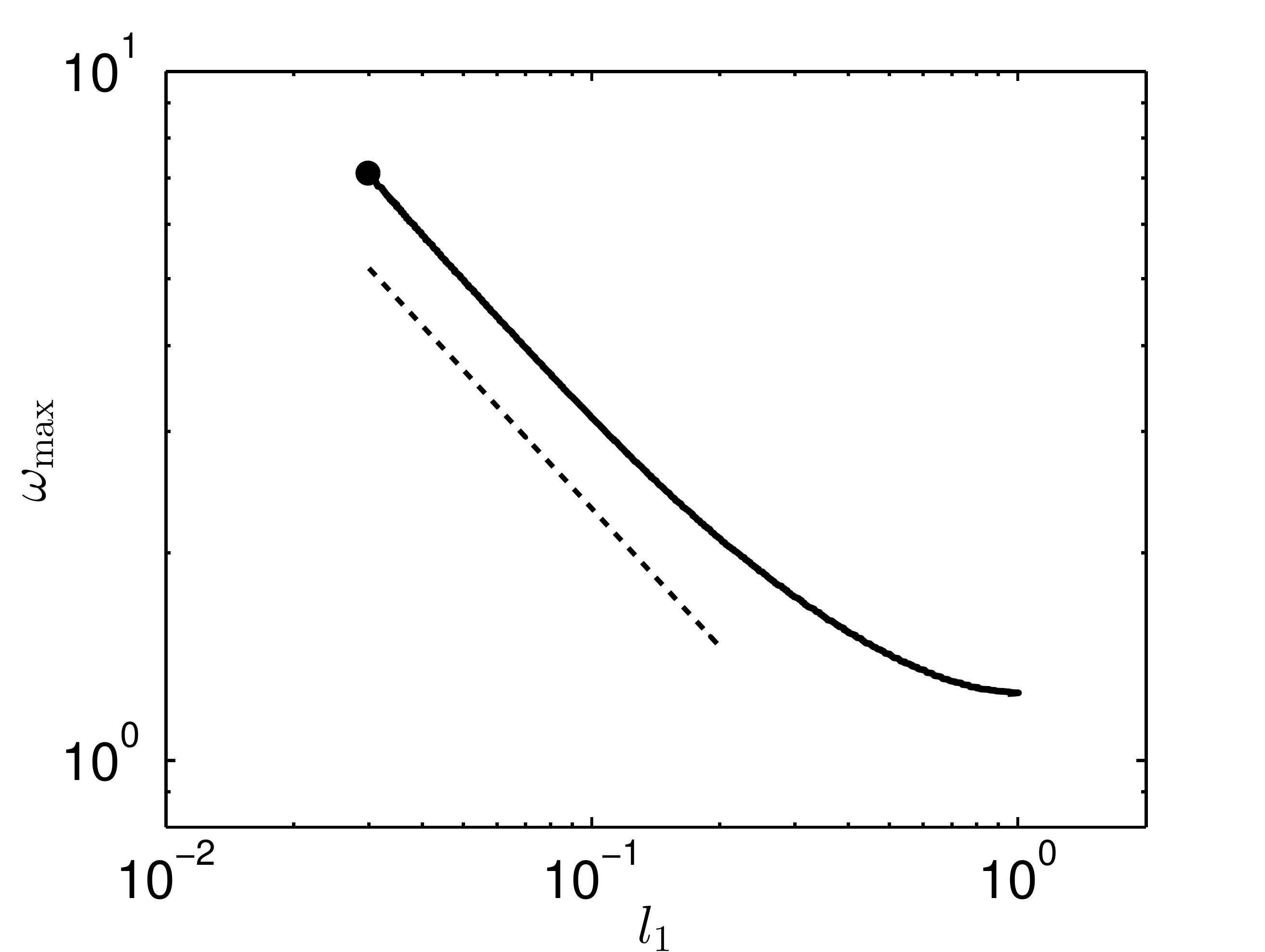}

\caption{The vorticity maximum $\omega_{\max}$ vs. the pancake thickness $l_{1}$ in logarithmic scales during the pancake evolution. The circle marks the vorticity maximum at the final time $t=7.5$ and the dashed line indicates the power-law $\omega_{\max}\propto l_{1}^{-2/3}$.
}
\label{fig:fig4}
\end{figure}

According to definition given in Section~\ref{Sec:NumResults}, the eigenvalues of the Hessian $\boldsymbol{\Gamma}^{(x)}$ are connected with the characteristic scales of the low-Jacobian structure as $\lambda_{i}^{(J)}=2J_{\min}/\ell_{i}^{2}$.
Then, the relation~(\ref{HessianJ_xaB}) can be rewritten as 
\begin{equation}\label{23relation-1}
\gamma_{ii}\approx 2J_{\min}\sigma_{i}^{2}/\ell_{i}^{2},\quad i=1,2,3.
\end{equation}
From the numerical simulation combined with the properties of the pancake model solution, we deduce that $\sigma_{1}\propto\ell_{1}\propto l_{1}$ and $\sigma_{2}\sim \ell_{2}\sim 1$, see Eqs.~(\ref{solution-exp-l}),~(\ref{solution-exp}),~(\ref{solution-S}). 
This means that the components $\gamma_{11}$ and $\gamma_{22}$ should decay as the Jacobian, $J_{\min}\propto e^{-\beta_{2}t}$. 
As shown in Fig.~\ref{fig:fig3}(d), these components do indeed decay, however they do not follow the mentioned dependency exactly, what may be connected with the difference between $\mathbf{U}$ and the unity matrix we observe in the simulations. 
For the large $(3,3)$-component we have $\ell_{3}\sim 1$ and $\sigma_{3}\propto e^{\beta_{3}t}\propto \omega_{\max}^{-1}l_{1}^{-1}$, so that 
\begin{equation}\label{23relation-2}
\gamma_{33}\approx 2J_{\min}\sigma_{3}^{2}/\ell_{3}^{2} \propto J_{\min}\omega_{\max}^{-2} l_{1}^{-2} \propto \omega_{\max}^{-3} l_{1}^{-2}, 
\end{equation}
where we used the inverse-proportionality between the vorticity maximum and the Jacobian minimum~(\ref{vorticity-maximal}). 
Hence, when the $(3,3)$-component does not depend on time significantly, see Fig.~\ref{fig:fig3}(d), the relation~(\ref{23relation-2}) leads to the $2/3$-scaling between the vorticity maximum and the pancake thickness during the pancake evolution, 
\begin{equation}\label{23relation-3}
\omega_{\max}\propto \gamma_{33}^{-1/3}l_{1}^{-2/3}, 
\end{equation}
as confirmed numerically in Fig.~\ref{fig:fig4}. 
Note that the $2/3$-scaling is observed for $\omega_{\max}\ge 2$, what corresponds to time interval $t\ge 4$, see Fig.~\ref{fig:fig1}(a). 
In this interval, the $(3,3)$-component deviates from its average value by no more than 20\%, Fig.~\ref{fig:fig3}(d), what leads to just 7\% deviation in $\gamma_{33}^{-1/3}$ standing in the relation~(\ref{23relation-3}). 

Thus, the $2/3$-scaling comes from the finiteness of the Hessian component $\gamma_{33}\propto 1$ and the lateral size $\ell_{3}\propto 1$ of the low-Jacobian structure. 
The finiteness of the Hessian elements $\gamma_{ij}$ can only come from the next-order corrections to the pancake model~(\ref{solution-1})-(\ref{solution-2}), since for the model these elements are zeroth, $\boldsymbol{\gamma}=\mathbf{0}$. 
Note that we saw manifestation of the next-order corrections already in~\cite{agafontsev2016asymptotic}, in the form of the pancake vorticity structure deviating from pure plane much larger than the pancake thickness, see e.g. Fig.~\ref{fig:fig1}(c). 
The latter meant that the model solution did not describe the whole pancake structure, however, we demonstrated that this model described locally an every nearly flat pancake segment with the model parameters gradually changing from one segment to another. 
In the $\mathbf{x}$-space, the main characteristic size of the pancake is its thickness, $l_{1}\ll 1$, and as we noted in Section~\ref{Sec:Pancake_model}, in the $\mathbf{a}$-space this size remains of unity order. 
We think that the gradual dependence of the pancake model parameters with the pancake segment together with the finiteness of the pancake in the $\mathbf{a}$-space are connected with the finiteness of the Hessian elements $\gamma_{ij}$, however, more study is necessary to clarify this phenomenon in details. 

The results discussed above are related to the high-vorticity structure corresponding to the global vorticity maximum and are obtained from the simulation of the initial condition $I_{2}$. 
We verified that several other pancakes from the same simulation corresponding to other local maxima of vorticity, as well as pancakes from the simulation of the initial flow $I_{1}$ discussed in Appendix~\ref{Sec:AppB}, demonstrate the same properties for the Jacobi and the Hessian matrices and follow the $2/3$-scaling~(\ref{introEq}). 

%----------------------------------------------------------------------------
%----------------------------------------------------------------------------

\section{Conclusions}
\label{Sec:Conclusions}

In the present paper we have studied high-vorticity structures developing in the incompressible 3D Euler equations in terms of the vortex line representation (VLR). 
The VLR is the transformation from the Eulerian coordinates of the flow to the Lagrangian markers of the vortex lines, and represents a partial integration of the Euler equations with respect to conservation of the Cauchy invariants. 
The latter means that a numerical simulation of the VLR equations must conserve the Cauchy invariants along the vortex line trajectories with the round-off accuracy. 
This property may be very important in the sense of the accuracy and control of the 3D Euler simulations while approaching sharp gradients. 
We have developed a new numerical method for the Euler equations in terms of the VLR and performed high-resolution simulations for two initial flows. 

As the first result, we have demonstrated that the growth of vorticity is related to the smallness of the Jacobian of the VLR, with the inverse-proportional relation between the two, $\omega_{\max}\propto 1/J_{\min}$. 
This agrees with the pancake model solution of~\cite{agafontsev2016asymptotic} and the relation~(\ref{omega-J}) derived under the assumption of unidirectional vorticity. 
Thus, a high-vorticity region for which the vorticity direction changes sharply with the coordinate may feature a different relation between the vorticity and the Jacobian. 
The pancake model solution turns out to be degenerate in terms of the VLR, so that all time-dependency for the vorticity comes from the denominator of Eq.~(\ref{Cauchy}), i.e., the Jacobian. 
The inverse of the Jacobian has the meaning of density of vortex lines, so that the vorticity within the pancake grows proportionally to this density. 
The latter is the manifestation of the compressibility of the vortex lines. 

A developing pancake structure affects the VLR mapping, which we examine with the singular value decomposition (SVD) of the Jacobi matrix. 
As indicated by the singular values, the mapping is strongly compressed along one direction with the rate proportional to the pancake thickness $\sigma_{1}\propto l_{1}\propto e^{-\beta_{1}t}$. 
Assuming that such behavior persists in the limit $t\to +\infty$, this may be seen as touching of the vortex lines, with the vorticity growing unboundly, $\omega_{\max}(t)\to +\infty$. 
Along the other two directions, the mapping either does not change substantially, $\sigma_{2}\propto 1$, or stretches as $\sigma_{3}\propto \omega_{\max}^{-1}l_{1}^{-1}\propto e^{\beta_{3}t}$. 
In the local coordinate system of the pancake, the rotation matrix of the SVD in the $\mathbf{x}$-space is close to unity, $\mathbf{U} \simeq \mathbf{1}$, while the rotation matrix in the $\mathbf{a}$-space $\mathbf{V}$ approaches to a constant matrix that depends on the initial parameters. 
The main characteristic size of the pancake -- its exponentially decaying thickness -- corresponds in the $\mathbf{a}$-space to distance of unity order. 
Thus, in the space of Lagrangian markers $\mathbf{a}$ the high-vorticity region does not shrink, in contrast to the previously made assumption~\cite{kuznetsov2000collapse} made by analogy with the gas dynamics case. 
These results also follow analytically from the VLR written for the pancake model solution of~\cite{agafontsev2016asymptotic}, what confirms the applicability of this model. 

In simulations, the Jacobian changes sharply along the pancake perpendicular direction $x_{1}$. 
This property can only come from the next-order corrections to the pancake model solution, since the Jacobian for the model does not depend on spatial coordinates. 
Assuming that these corrections are present, we demonstrate that the Hessian $\boldsymbol{\gamma}$ for the Jacobian in the basis induced by rotation matrix $\mathbf{V}$ must be close to diagonal and the sharp dependency for the Jacobian along the $x_{1}$-axis comes from small but finite element $\gamma_{11}$. 

The pancake model solution allows for an arbitrary power-law scaling between the vorticity maximum and the pancake thickness, given by the ratio of the exponents $\beta_{2}/\beta_{1}$. 
For the first time, we discovered numerically that this ratio is close to $2/3$ in~\cite{agafontsev2015}, however, we were not able to explain this observation. 
With the present VLR study, we identify that the $2/3$-scaling~(\ref{introEq}) comes from the finite Hessian element $\gamma_{33}\propto 1$ and the finite lateral pancake size $\ell_{3}\propto 1$. 
We think that the finiteness of $\gamma_{33}$ is connected with the two properties of the pancake structures, namely, the gradual dependence of the pancake model parameters with the pancake segment and the finiteness of the pancake thickness in the $\mathbf{a}$-space. 
However, more study is necessary to clarify this connection in details. 

Our approach utilizes the general properties of the frozen-in-fluid fields and, potentially, can be generalized to a wider group of physical phenomena far beyond the scope of this paper. 
For instance, compressibility of magnetic field lines~\cite{kuznetsov2004compressible} should play an essential role in generation of magnetic filaments in the convective zone of the Sun and in the magnetic dynamo theories in space plasma, see e.g.~\cite{moffatt1978field,childress2008stretch}. 
As shown in~\cite{kuznetsov2007effects,kudryavtsev2013statistical,kuznetsov2015anisotropic}, compressible character of the frozen-in-fluid divorticity field is an important factor in the formation of direct Kraichnan cascade in 2D hydrodynamic turbulence.

\section*{Acknowledgements}

The work of D.S.A. and E.A.K. was supported by the Russian Science Foundation (grant 14-22-00174). 
D.S.A. acknowledges the support from IMPA during the visits to Brazil. 
A.A.M. was supported by the RFBR grant 17-01-00622, the CNPq (grant 302351/2015-9) and the Program FAPERJ Pensa Rio (grant E-26/210.874/2014). 
The simulations were performed at the Novosibirsk Supercomputer Center (NSU) and the analysis of the results was done at the Data Center of IMPA (Rio de Janeiro).

\newpage
\appendix

%----------------------------------------------------------------------------
%----------------------------------------------------------------------------

\section{Numerical scheme}
\label{Sec:App0}

The VLR system of equations~(\ref{a-t})--(\ref{v-normal-v}) rewritten for the periodic mapping~(\ref{b-functions}) is solved numerically in the periodic box $\mathbf{x}=(x_1,x_2,x_3)\in[-\pi ,\pi]^{3}$ with the Runge--Kutta forth-order pseudo-spectral method. 
To avoid the so-called bottle-neck instability, at each time step we perform filtering in the Fourier space with the cut-off function~\cite{hou2007computing} 
\begin{equation}\label{dumping}
\rho(\mathbf{k})=\exp\bigg(-36\sum_{j}(k_{j}/K_{\max}^{(j)})^{36}\bigg),
\end{equation}
which cuts approximately 20\% of modes at the edges of the spectral band in each direction. 
Here $K_{\max}^{(j)} = N_{j}/2$ are the maximal wavenumbers and $N_{j}$ are numbers of nodes along directions $j=1,2,3$. 
Additionally, we performed simulations using different Fourier filters $\rho(\mathbf{k})$ including the standard dealiasing rule, and found the results practically identical. 
The adaptive time stepping is implemented via the CFL stability criterion with the Courant number $0.5$. 
The adaptive anisotropic rectangular grid is uniform for each direction and adapted independently along each spatial coordinate. 
The idea for the adaption comes from the standard dealiasing rule. 
At early times, the Fourier spectrum of the solution is concentrated at low harmonics, while higher harmonics contain numerical noise. 
We track the ``signal-noise'' boundary~\cite{agafontsev2015} until it reaches $K_{\max}^{(j)}/2$ for any of the three directions $j=1,2,3$, and then refine the grid along the corresponding direction. 
This rule is optimized for cubic nonlinearity of the VLR equations~(\ref{a-t})--(\ref{v-normal-v}), while for the direct Euler simulations with quadratic nonlinearity we used $2K_{\max}^{(j)}/3$, see~\cite{agafontsev2015,agafontsev2016development,agafontsev2016asymptotic}. 
Transition to a refined grid is performed with the Fourier interpolation, which has an error comparable with the round-off. 
While the simulation is running in this way, the aliasing error is avoided and the cut-off function~(\ref{dumping}) affects only harmonics containing numerical noise. 

We start simulations in cubic grid $192^{3}$ and refine the grid until the total number of nodes $N_{1}N_{2}N_{3}$ reaches $1536^{3}$, then continuing with the fixed grid. 
As a stopping criterion, at $K_{\max}^{(j)}/2$ we compare the Fourier spectrum of the mapping~(\ref{b-functions}) along each spatial direction with $10^{-13}$ times its maximum value, see~\cite{agafontsev2015} where the analogous procedure was implemented. 
After the grid is fixed, aliasing in Eq.~(\ref{omega-r}) becomes significant, but it contributes mainly to harmonics close to the edges of the spectral band.
To diminish this contribution, we optimize the shape of the cut-off function dynamically for the vorticity function, which features a significantly narrower Fourier spectrum. 

The simulation of $I_{2}$ initial flow limited by $1536^3$ total number of nodes reaches the final time $t=7.5$ in the grid $1458\times 648\times 3456$. 
We compared its results with two simulations in grids of $512^3$ and $1024^3$ nodes finished at $t=6.02$ and $7.16$, respectively. 
The maximal relative point-by-point difference for the mapping $|\mathbf{b}^{(1)}(\mathbf{x})-\mathbf{b}^{(2)}(\mathbf{x})|/|\mathbf{b}^{(1)}(\mathbf{x})|$ between any two of these simulations does not exceed $10^{-5}$ order up to the earliest final time. 
Simulations performed with twice less time steps finish with the same final time and grid, and yield relative errors for the mapping of $10^{-9}$ order maximum. 
We obtain analogous results for the convergence study with the $I_{1}$ initial flow as well. 
These tests confirm that the errors in calculation of the vortex line trajectories in Eq.~(\ref{a-t}) are very small and do not affect our results. 

We also compared the VLR simulations with the direct simulations of the Euler equations~(\ref{Euler2}) performed as described in~\cite{agafontsev2015,agafontsev2016asymptotic} in grids limited by $2048^{3}$ nodes; for $I_{2}$ initial flow the direct simulation ended at $t=8.92$ with the final grid $1944\times 972\times 4374$. 
The maximal relative point-by-point difference for the vorticity field $|\omegabold^{(1)}(\mathbf{x})-\omegabold^{(2)}(\mathbf{x})|/|\omegabold^{(1)}(\mathbf{x})|$ between the VLR and the direct simulations is kept below $10^{-9}$ before the grid of the VLR simulation is fixed at $t=5.39$, and then increases up to $3\times 10^{-5}$ at the final time $t=7.5$. 
Both the VLR and the direct numerical schemes conserve the total energy $E=(1/2)\int\mathbf{v}^2\,d^{3}\mathbf{x}$ and the helicity $\Omega=\int (\mathbf{v}\cdot\omegabold)\,d^{3}\mathbf{x}$ with a relative error smaller than $10^{-11}$. 
This provides independent verification for the accuracy of both numerical schemes. 

%----------------------------------------------------------------------------
%----------------------------------------------------------------------------

\section{Initial conditions for numerical simulations}
\label{Sec:AppA}
\renewcommand{\arraystretch}{0.5}

We consider initial vorticity at $t = 0$ in the form of Fourier series 
\begin{equation}
\omegabold_0(\mathbf{x}) =  \sum_{\mathbf{h}} 
\left[\mathbf{A}_\mathbf{h}\cos(\mathbf{h}\cdot\mathbf{x})
+\mathbf{B}_\mathbf{h}\sin(\mathbf{h}\cdot\mathbf{x})\right],
\label{IC1}
\end{equation}
where $\mathbf{h} = (h_1,h_2,h_3)$ is a vector with integer components. 
Due to incompressibility, real vectors $\mathbf{A}_\mathbf{h}$ and $\mathbf{B}_\mathbf{h}$ must satisfy orthogonality conditions, $\mathbf{h}\cdot\mathbf{A}_\mathbf{h} = \mathbf{h}\cdot\mathbf{B}_\mathbf{h} = 0$, necessary for self-consistency. 
The paper is based on two selected initial conditions $I_{1}$ and $I_{2}$, for which we provide all nonzero vectors $\mathbf{A}_\mathbf{h}$ and $\mathbf{B}_\mathbf{h}$ in the Tables below. 
Note that the number of excited harmonics in the initial vorticity affects directly the speed of the simulation, because in Eq.~(\ref{omega-r}) we need to calculate $\omegabold_{0}(\mathbf{a})$ for a ``shifted'' $\mathbf{a}$-grid at each time step. 

\begin{table}[H]
\caption{Nonzero coefficients in Eq.~(\ref{IC1}) for the initial vorticity field $I_{1}$ with the average energy density $E/(2\pi)^{3}\approx 0.54$ and the average helicity density $\Omega/(2\pi)^{3}\approx 1.05$.}
\begin{center}
 \begin{tabular}{| c | c | c |}
  \hline
  $\mathbf{h}$ & $\mathbf{A_{h}}$ & $\mathbf{B_{h}}$ \\ \hline
  (-1,0,2) & (0.0065641, 0.0027931, 0.003282) & (0.0044136, 0.0056271, 0.0022068) \\ \hline
  (0,0,0) & (0.065101, 0.0005801, -0.064109) & (0.0045744, -0.022895, 0.18392) \\ \hline
  (0,0,1) & (0, 1, 0) & (1, 0, 0) \\ \hline
  (0,0,2) & (0, 0.01, 0) & (0.01, 0, 0) \\ \hline
  (0,1,0) & (0.21204, 0, -0.070625) & (-0.14438, 0, 0.23298) \\ \hline
  (0,1,1) & (0.045977, -0.010151, 0.010151) & (0.041942, 0.040326, -0.040326) \\ \hline
  (0,2,0) & (0.005, 0, 0) & (0, 0, 0.005) \\ \hline
  (1,0,0) & (0, 0, 0.1) & (0, 0.1, 0) \\ \hline
  (1,0,1) & (-0.046112, 0.017081, 0.046112) & (-0.0097784, 0.020122, 0.0097784) \\ \hline
  (1,1,2) & (-0.0034664, 0.0049556, -0.00074462) & (-0.0059316, -0.0010472, 0.0034894) \\ \hline
  (2,0,0) & (0, 0, 0.02) & (0, 0.02, 0) \\ \hline
 \end{tabular}
\end{center}
 \label{tab:M1}
\end{table}

\begin{table}[H]\footnotesize
\caption{Nonzero coefficients in Eq.~(\ref{IC1}) for the initial vorticity field $I_{2}$ with the average energy density 
$E/(2\pi)^{3}\approx 0.55$ and the average helicity density $\Omega/(2\pi)^{3}\approx 1$.}
\begin{center}
 \begin{tabular}{| c | c | c |}
  \hline
  $\mathbf{h}$ & $\mathbf{A_{h}}$ & $\mathbf{B_{h}}$ \\ \hline
  (-1,0,1) & (-0.040618, 0.039651, -0.040618) & (-0.030318, 0.064657, -0.030318) \\ \hline
  (0,0,0) & (0.067751, -0.1311, -0.11256) & (-0.082614, -0.0364, 0.18932) \\ \hline
  (0,0,1) & (0, 1, 0) & (1, 0, 0) \\ \hline
  (0,1,1) & (0.0062549, 0.044315, -0.044315) & (0.034983, -0.014521, 0.014521) \\ \hline
  (1,0,0) & (0, 0.079395, 0.07027) & (0, 0.099411, 0.012762) \\ \hline
  (1,1,0) & (-0.047174, 0.047174, -0.045572) & (-0.049622, 0.049622, 0.001773) \\ \hline
 \end{tabular}
\end{center}
 \label{tab:M2}
\end{table}

%----------------------------------------------------------------------------
%----------------------------------------------------------------------------

\section{Simulation results for the $I_{1}$ initial flow}
\label{Sec:AppB}

\begin{figure}[H]
\centering
\includegraphics[width=7.5cm]{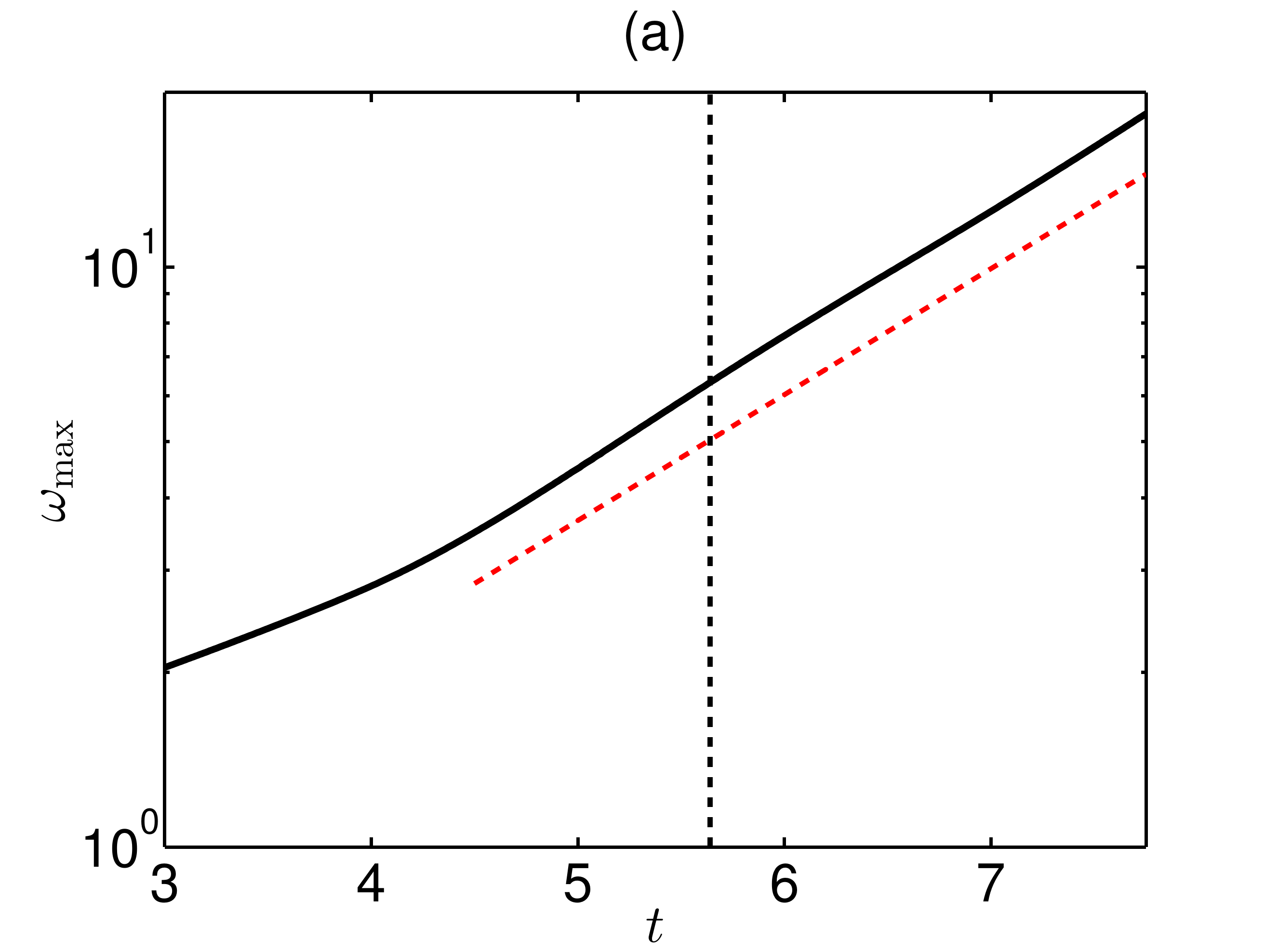}
\includegraphics[width=7.5cm]{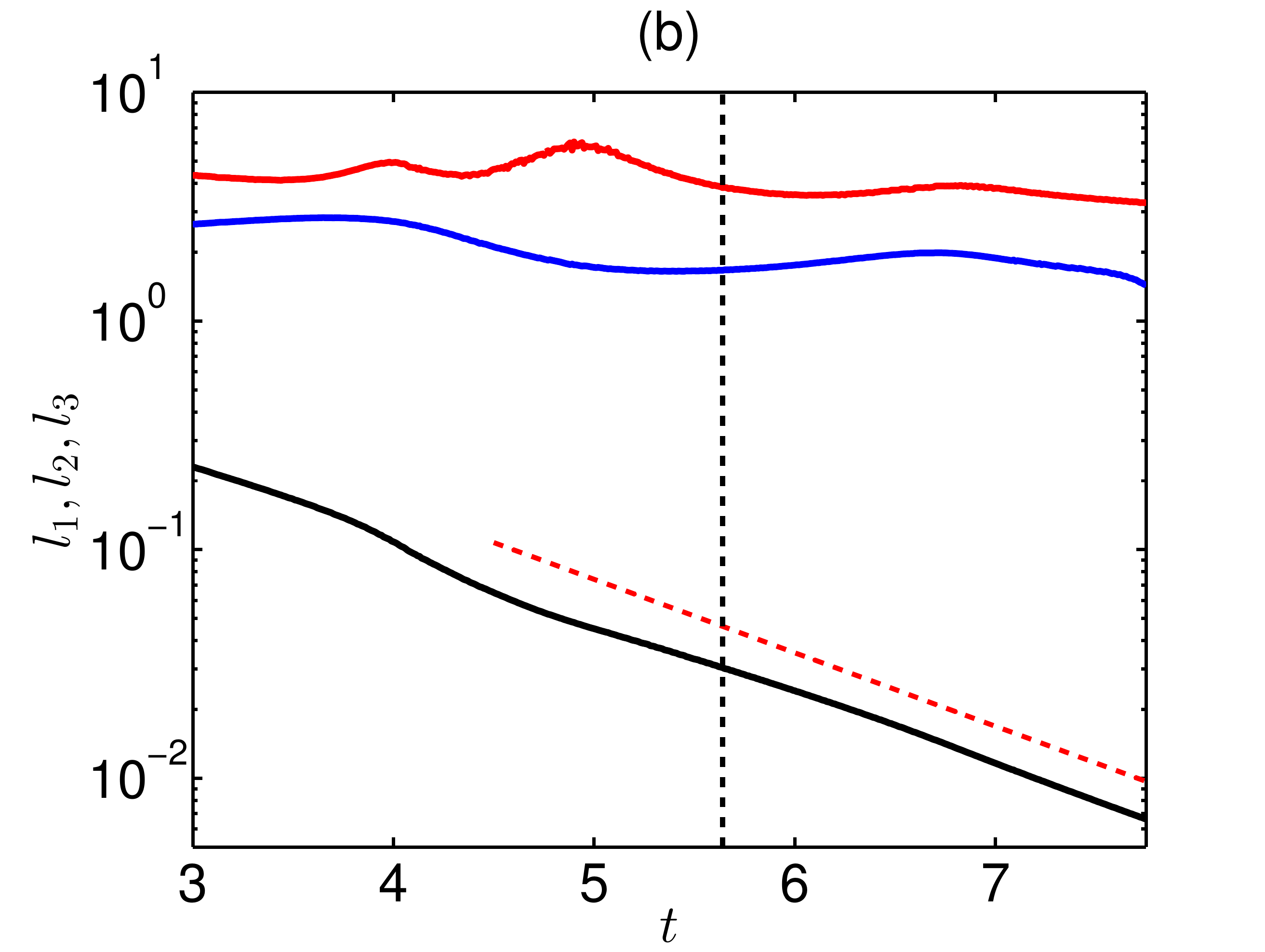}
\includegraphics[width=7.5cm]{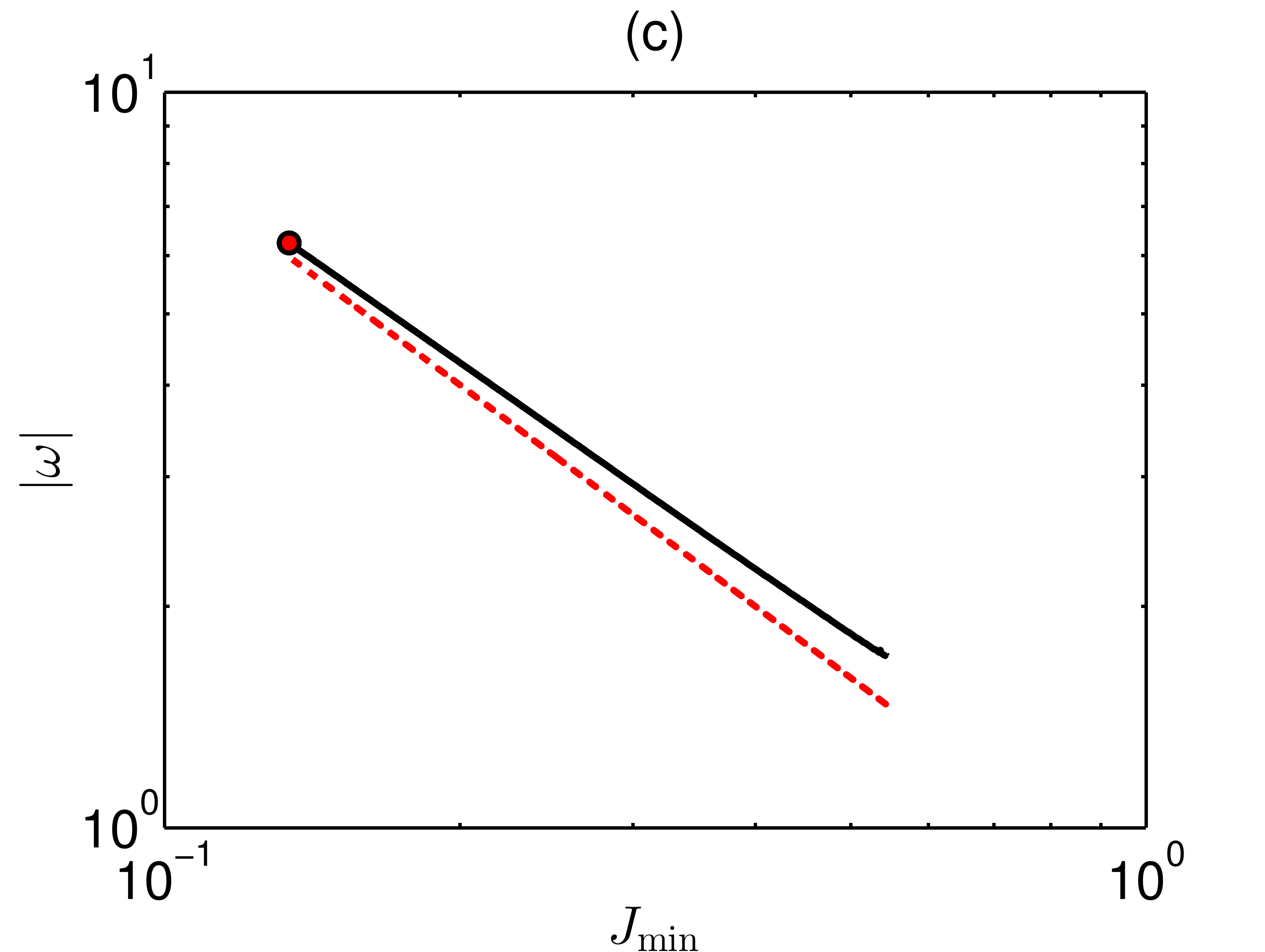}
\includegraphics[width=7.5cm]{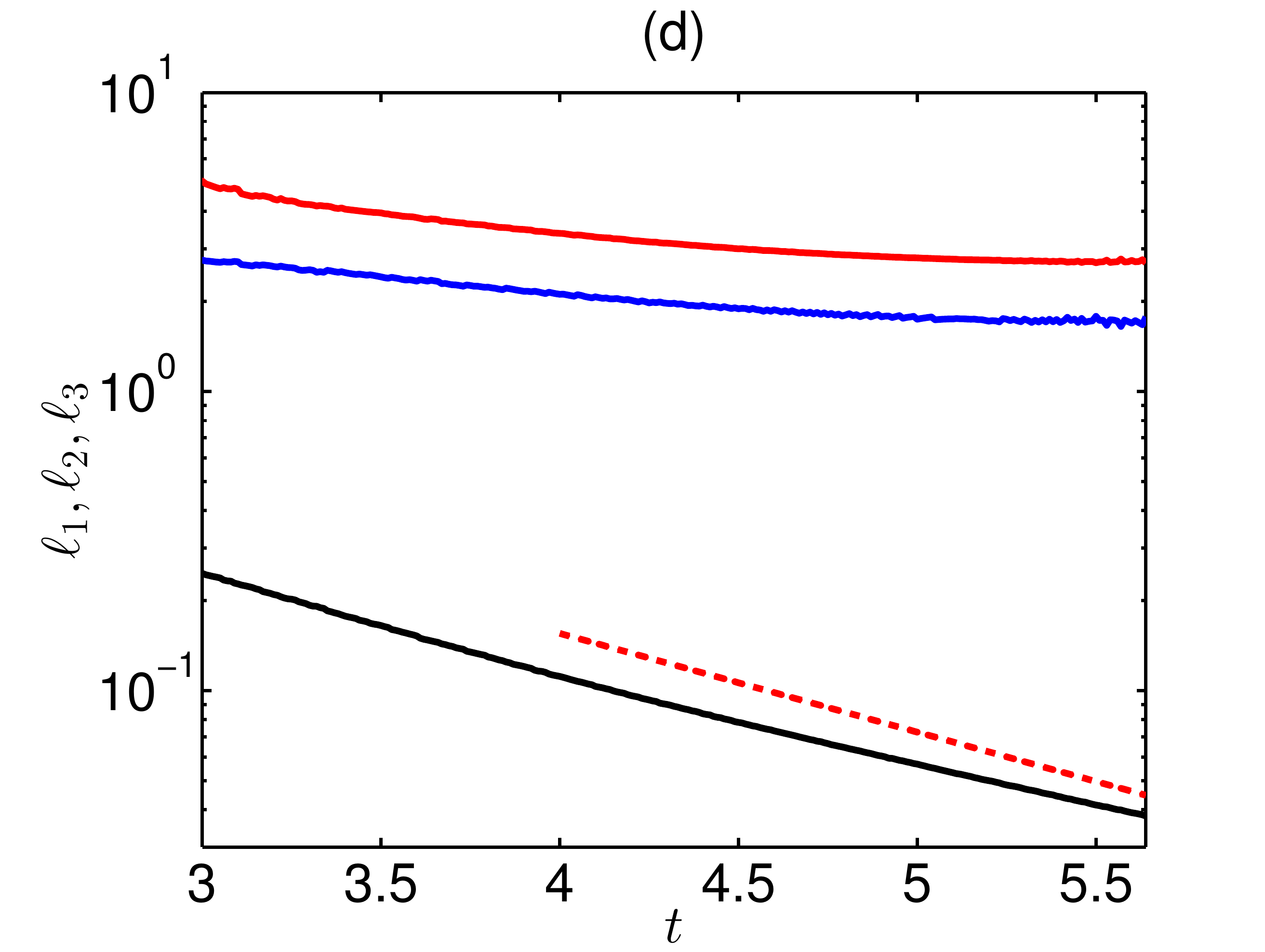}
\caption{\textit{(Color on-line)} 
(a) Global vorticity maximum as a function of time (logarithmic vertical scale) for the direct simulation of the Euler equations~(\ref{Euler2}). The red dashed line indicates the slope $\propto e^{\beta_{2}t}$ with $\beta_{2}=0.5$. The vertical dashed line marks the final time $t = 5.64$ for the simulation of the VLR equations~(\ref{a-t})--(\ref{v-normal-v}). 
(b) Time-evolution of the characteristic spatial scales $l_{1}$ (black), $l_{2}$ (blue) and $l_{3}$ (red) for the respective pancake vorticity structure. The red dashed line indicates the slope $\propto e^{-\beta_{1}t}$ with $\beta_{1}=0.74$. 
(c) Relation between the vorticity and the Jacobian at the Jacobian minimum $J_{\min}$ closest to the global vorticity maximum, the VLR simulation. The red dashed line indicates the asymptotic relation $\omega \propto 1/J_{\min}$, while the circle marks the final time $t=5.64$. 
(d) Time-evolution of the characteristic spatial scales $\ell_{1}$ (black), $\ell_{2}$ (blue) and $\ell_{3}$ (red) for the low-Jacobian structure corresponding to the minimum $J_{\min}$. The black dashed line indicates the slope $\propto e^{-\beta_{1}t}$ with $\beta_{1}=0.74$. 
}
\label{fig:AppB-fig1}
\end{figure}

In this Appendix we provide some results for the simulation of the $I_{1}$ initial condition from Tab.~\ref{tab:M1}. 
The direct simulation of the Euler equations~(\ref{Euler2}) for this flow in $2048^{3}$ grid is given in~\cite{agafontsev2016asymptotic}. 
It ends at $t=7.75$ with the grid $972\times 2048\times 4096$ and the global vorticity maximum increased from $1.5$ at $t=0$ to $18.4$ at the final time. 
At late times, this growth is exponential, $\omega_{\max}\propto e^{\beta_{2}t}$ with $\beta_{2}=0.5$, while the pancake thickness decays as $l_{1}\propto e^{-\beta_{1}t}$, $\beta_{1}=0.74$, and the lateral pancake dimensions do not change significantly, $l_{2,3}\propto 1$, see Fig.~\ref{fig:AppB-fig1}(a,b). 

\begin{figure}[t]
\centering
\includegraphics[width=7.5cm]{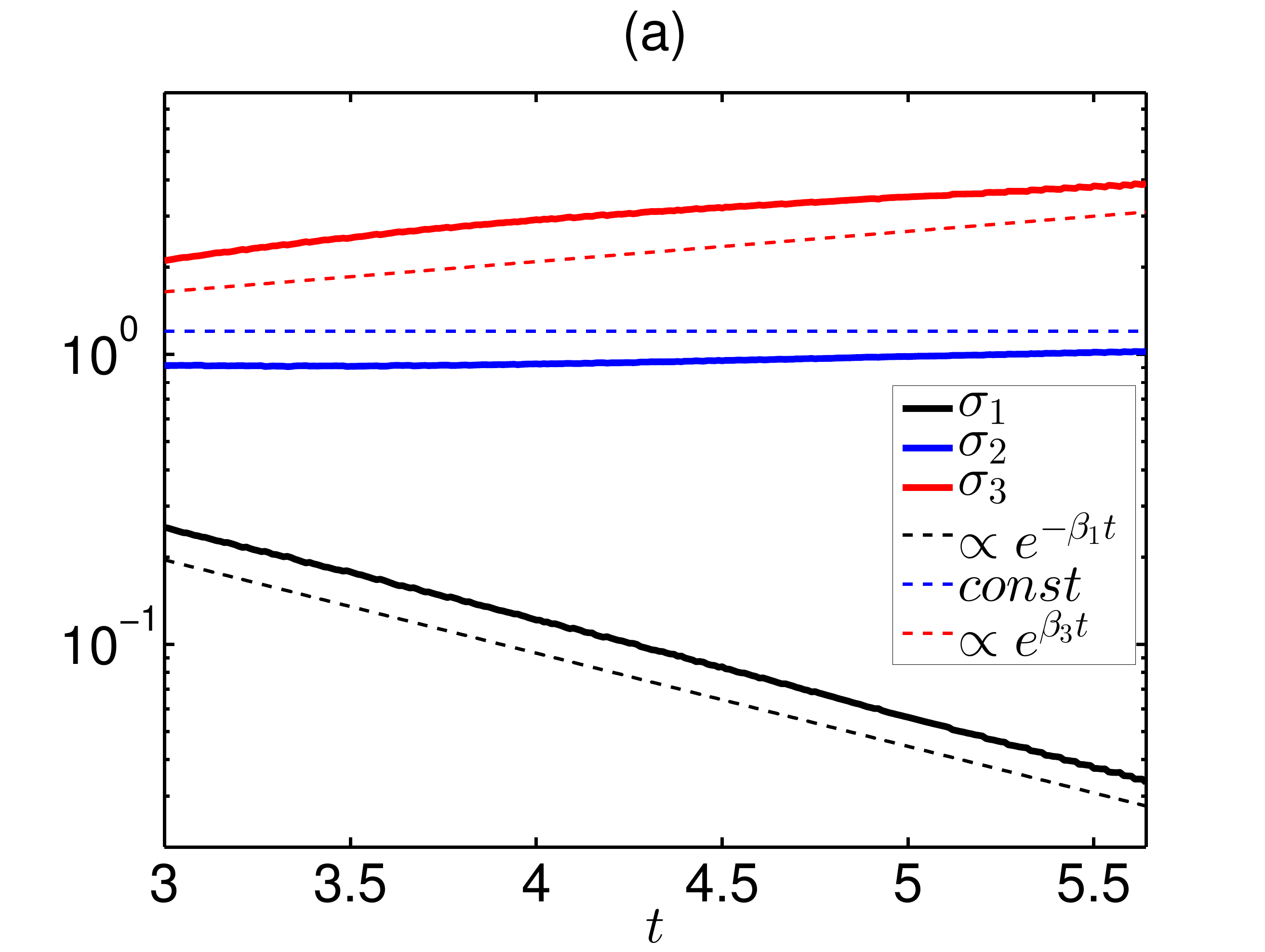}
\includegraphics[width=7.5cm]{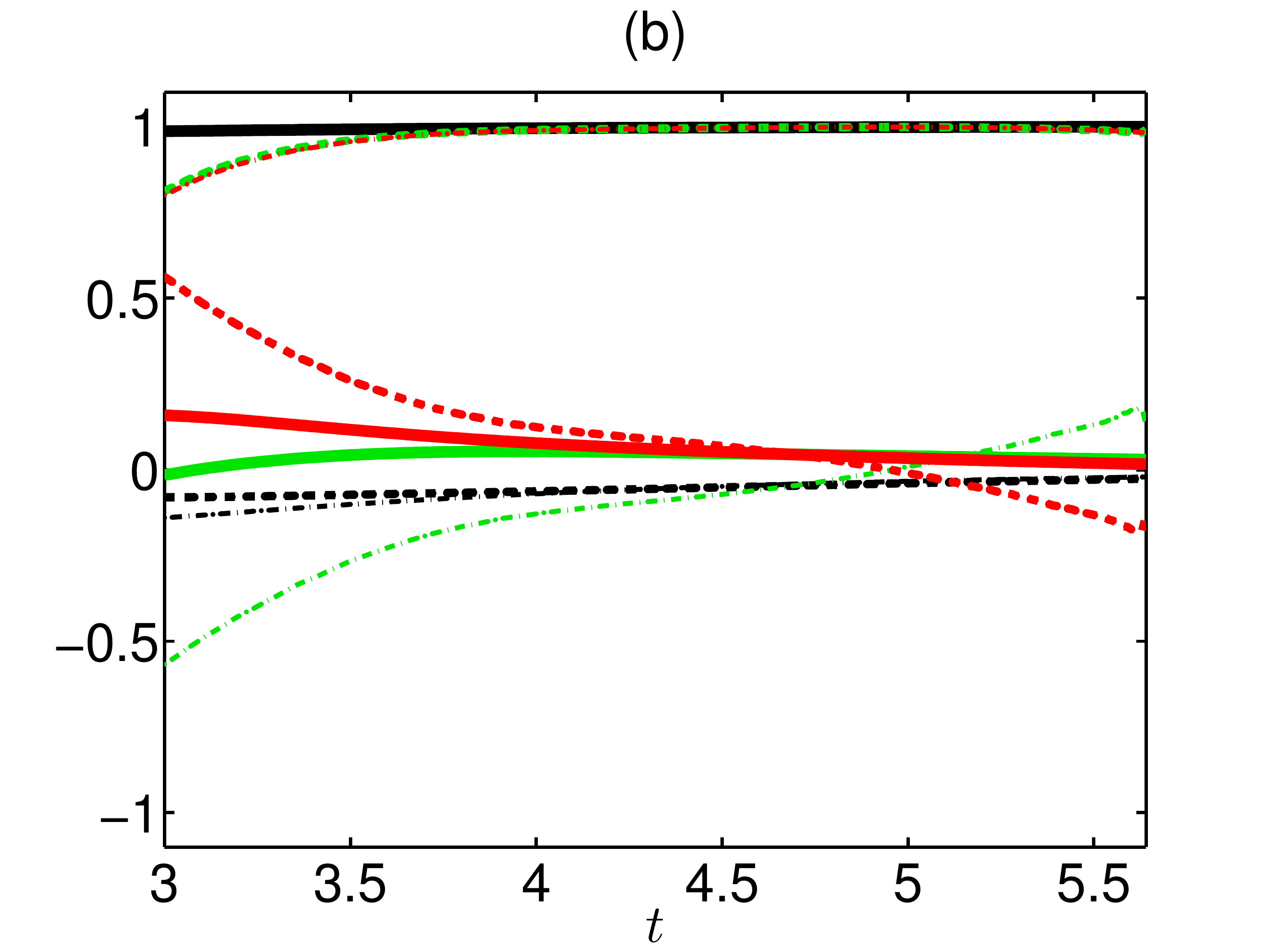}
\includegraphics[width=7.5cm]{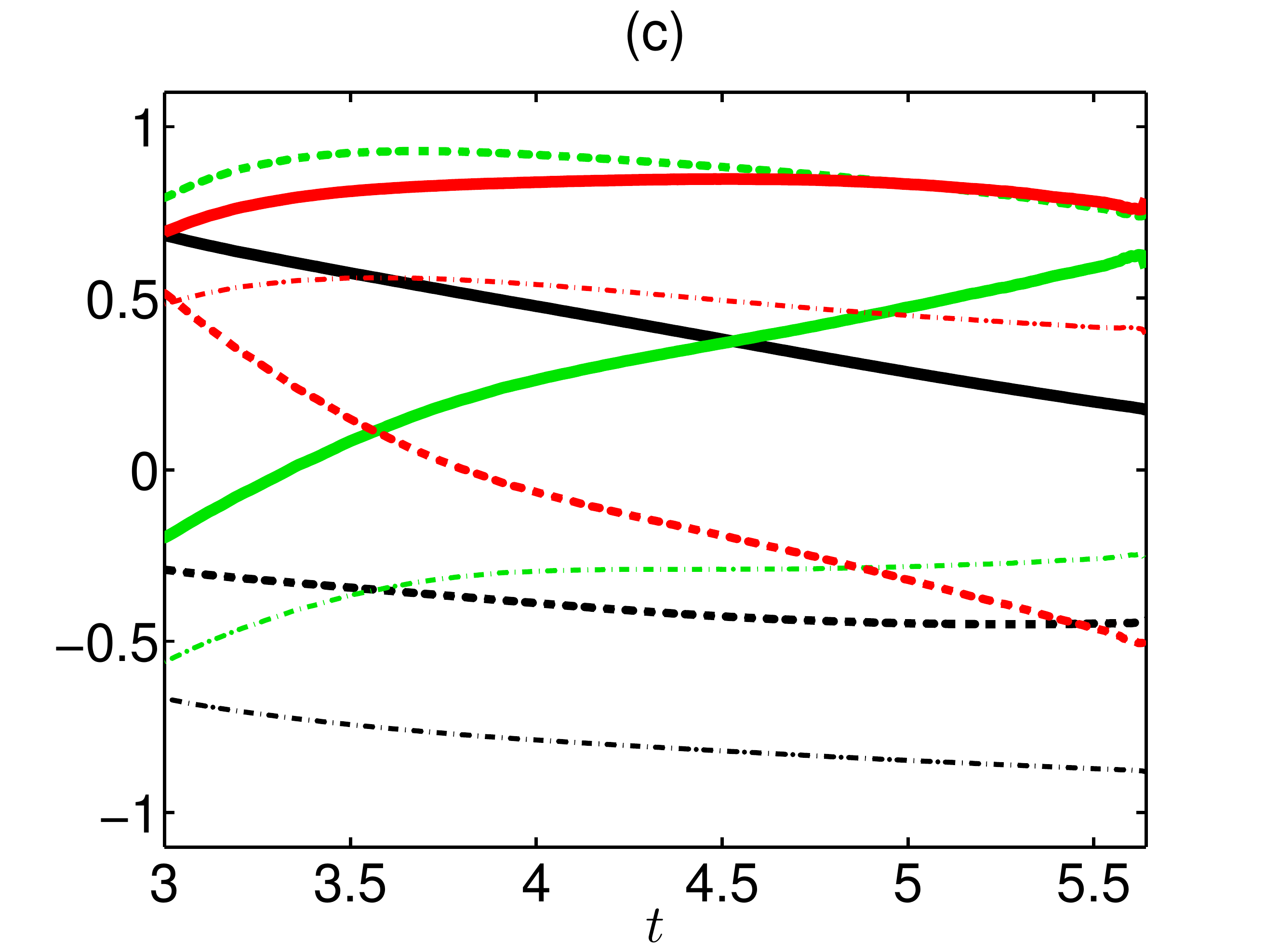}
\includegraphics[width=7.5cm]{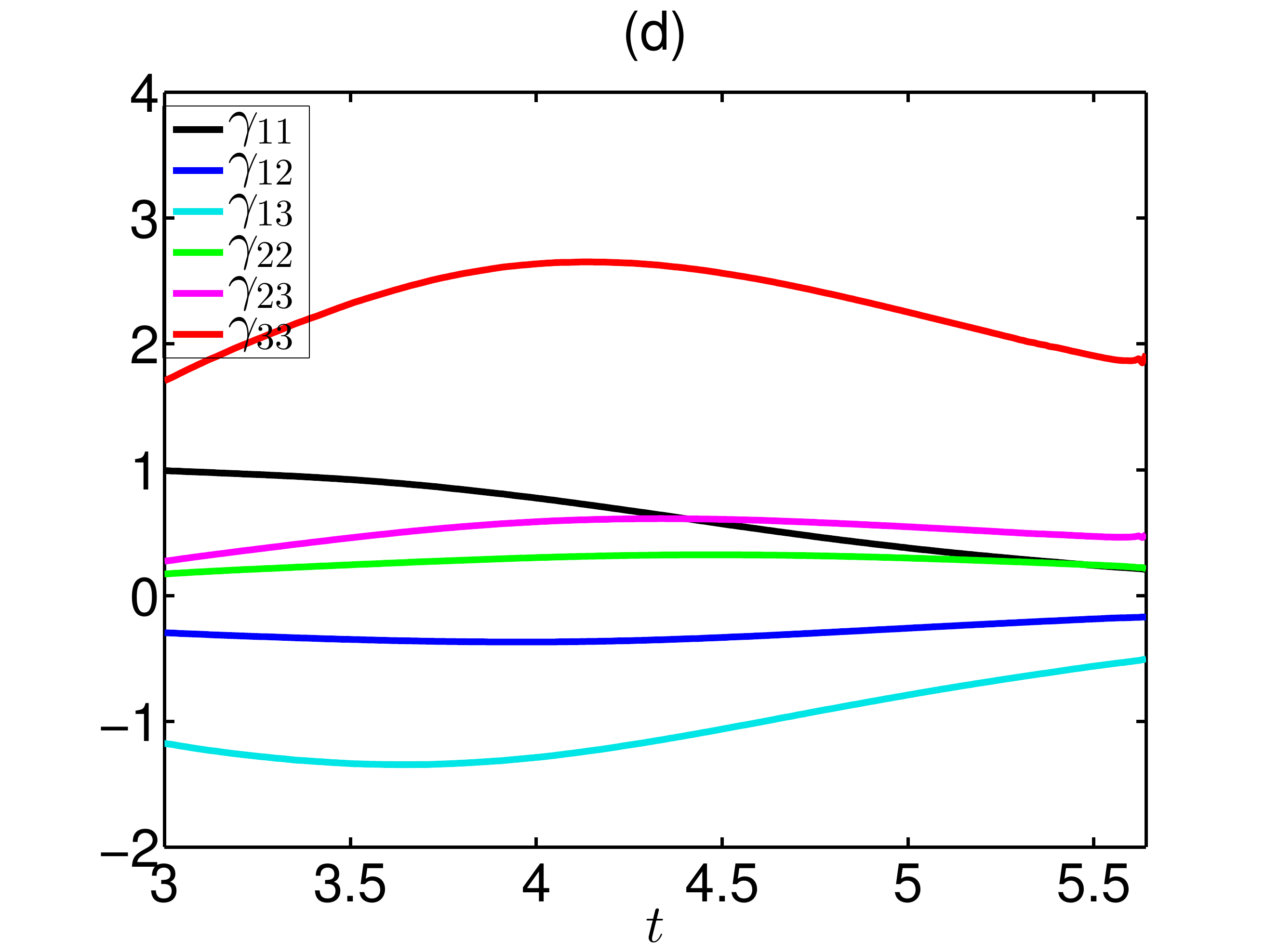}
\caption{\textit{(Color on-line)} 
(a) Singular values $\sigma_{1}$ (black), $\sigma_{2}$ (blue) and $\sigma_{3}$ (red) of the Jacobi matrix $\widehat{\mathbf{J}}$ computed at $J_{\min}$, as functions of time. Dashed lines show the exponential slopes~(\ref{solution-S}) with $\beta_{1}=0.74$, $\beta_{2}=0.5$ and $\beta_{3}=0.24$. 
(b) Components of the rotation matrix $\mathbf{U}$ in $\mathbf{x}$-space, as functions of time. Thick black curve shows the $(1,1)$ component, dashed black -- $(1,2)$, thin dash-dot black -- $(1,3)$, thick green -- $(2,1)$, dashed green -- $(2,2)$, thin dash-dot green -- $(2,3)$, thick red -- $(3,1)$, dashed red -- $(3,2)$, thin dash-dot red -- $(3,3)$. 
(c) Same for the rotation matrix $\mathbf{V}$ in $\mathbf{a}$-space. 
(d) Evolution of the Hessian $\boldsymbol{\gamma} = \mathbf{V}^{T}\boldsymbol{\Gamma}^{(a)}\mathbf{V}$ in $\hat{V}$-basis. 
}
\label{fig:AppB-fig3}
\end{figure}

The VLR simulation ends at $t=5.64$ with the grid $648\times 1536\times 3456$ and the vorticity maximum $6.3$ at the final time, as indicated by the vertical line in Fig.~\ref{fig:AppB-fig1}(a). 
The maximal relative point-by-point difference for the vorticity field $|\omegabold^{(1)}(\mathbf{x})-\omegabold^{(2)}(\mathbf{x})|/|\omegabold^{(1)}(\mathbf{x})|$ between the VLR and the direct simulations is kept below $10^{-9}$ before the grid of the VLR simulation is fixed at $t=4.5$, and then increases up to $4\times 10^{-5}$ at the final time $t=5.64$. 
At late times, the Jacobian minimum close to the global vorticity maximum decays exponentially and inverse-proportionally to the vorticity maximum, $J_{\min}\propto e^{-\beta_{2}t}\propto\omega_{\max}^{-1}$, see Fig.~\ref{fig:AppB-fig1}(c). 
The thickness of the corresponding low-Jacobian region decays as the vorticity pancake thickness, $\ell_{1}\propto e^{-\beta_{1}t}\propto l_{1}$, with the longitudinal scales not changing substantially, $\ell_{2,3}\propto 1$, Fig.~\ref{fig:AppB-fig1}(d). 
The first singular value of the Jacobi matrix computed at $J_{\min}$ decays as the vorticity pancake thickness, $\sigma_{1}\propto e^{-\beta_{1}t}\propto l_{1}$, the second one does not change significantly, $\sigma_{2}\propto 1$, and the third singular value increases close to exponentially, Fig.~\ref{fig:AppB-fig3}(a). 
The rotation matrices of the SVD do not change substantially and matrix $\mathbf{U}$ remains close to unity, Fig.~\ref{fig:AppB-fig3}(b,c). 
The Hessian of the Jacobian $\boldsymbol{\gamma}$ is almost diagonal, has only the $(3,3)$-component of unity order which does not change significantly in time, while the other components are small, Fig.~\ref{fig:AppB-fig3}(d). 
Hence, we conclude that the simulation of $I_{1}$ initial flow leads to the same conclusions on the formation of intense vorticity and the emergence of the $2/3$-scaling from the geometrical properties of the VLR, as deduced from the $I_{2}$ simulation in the main text of this paper. 

\end{document}